\documentclass[12pt,preprint]{aastex}
\newcommand\eq{\begin{equation}}
\newcommand\eeq{\end{equation}}
\newcommand\eqn{\begin{eqnarray}}
\newcommand\eeqn{\end{eqnarray}}
\newcommand\kms{km s$^{-1}$}

\usepackage{ulem}
\slugcomment{AJ accepted Apr 07, 2009}

\shorttitle{O/H abundance distribution}
\shortauthors{Lagos et al.}

\begin{document}

\title{On the compact HII galaxy UM 408 as seen by GMOS-IFU: Physical conditions.}


\author{Patricio Lagos\altaffilmark{1,2}}
\email{plagos@iac.es}

\author{Eduardo Telles\altaffilmark{1,3}}
\email{etelles@on.br}

\author{Casiana Mu\~noz-Tu\~n\'on\altaffilmark{2}}
\email{cmt@iac.es}

\author{Eleazar R. Carrasco\altaffilmark{4}}
\email{rcarrasco@gemini.edu}

\author{Fran\c{c}ois Cuisinier\altaffilmark{5}}
\email{francois@ov.ufrj.br}

\and

\author{Guillermo Tenorio-Tagle\altaffilmark{6}}
\email{gtt@inaoep.mx}

\altaffiltext{1}{Observat\'{o}rio Nacional, Rua Jos\'{e} Cristino, 77, Rio de Janeiro,
20921-400, Brazil}
\altaffiltext{2}{Instituto de Astrofisica de Canarias, Via L\'actea s/n, 38200
La Laguna, Spain}
\altaffiltext{3}{Dept. of Astronomy, University of Virginia, P.O.BOX 400325, Charlottesville, VA, 22904-4325, USA}
\altaffiltext{4}{Gemini Observatory/AURA, Southern Operations Center,
Casilla 603,  La Serena, Chile}
\altaffiltext{5}{Observat\'orio do Valongo UFRJ, Ladeira do Pedro
Ant\^{o}nio 43, Rio de Janeiro, 20080-090, Brazil}
\altaffiltext{6}{Instituto Nacional de Astrofisica \'Optica y Electr\'onica,
AP 51, 72000 Puebla, M\'exico}

\begin{abstract}

We present Integral Field Unit GMOS-IFU data of the compact HII galaxy UM 408,
obtained at Gemini South telescope, in order to derive the spatial distribution
of emission lines and line ratios, kinematics, plasma parameters, and oxygen abundances as
well the integrated properties over an area of 3$\arcsec
\times$4$\arcsec$.4 equivalent with $\sim$750 $\times$ 1100 pc located in the central part of the galaxy. 
The starburst in this area is resolved into two giant regions of about 1$\arcsec$.5 and
1$\arcsec$ ($\sim$375 and $\sim$250 pc) diameter, respectively and separated 1.5-2$\arcsec$
($\sim$500 pc). The extinction distribution concentrate its highest values close
but not coincident with the maxima of H$\alpha$ emission around each one of the
detected regions. This indicates that the dust has been displaced from the
exciting clusters by the action of their stellar winds. The ages of these two
regions, estimated using H$\beta$ equivalent widths, suggest that they are
coeval events of $\sim$5 Myr with stellar masses of $\sim$10$^{4}$M$_{\odot}$. 
We have also used [OIII]/H$\beta$ and
[SII]/H$\alpha$ ratio maps to explore the excitation mechanisms in this galaxy.
Comparing the data points with theoretical diagnostic models, we found that all
of them are consistent with  excitation by photoionization by massive stars. The
H$\alpha$ emission line was used to measure the radial velocity and velocity
dispersion. The heliocentric radial velocity shows an apparent systemic motion
where the east part of the galaxy is blueshifted, while the west part is
redshifted, with a relative motion of $\sim 10$ km s$^{-1}$. The velocity
dispersion map shows supersonic values typical for extragalactic HII regions.
Oxygen abundances were calculated from the [OIII]$\lambda\lambda$4959,5007/[OIII]$\lambda$4363
ratios. We derived an integrated oxygen
abundance of 12+log(O/H)=7.87 summing over all spaxels in our field of view. 
An average value of 12+log(O/H)=7.77 and a difference of
$\Delta$(O/H)=0.47 between the minimum and maximum values
(7.58$\pm$0.06-8.05$\pm$0.04) were found, considering all data points where the
oxygen abundance was measured. The spatial distribution of oxygen abundance does
not show any significant gradient across the galaxy. On the other hand, the bulk
of data points are lying in a region of $\pm$2$\sigma$ dispersion (with
$\sigma$=0.1 dex) around the average value, confirming that this compact HII
galaxy as other previously studied dwarf irregular galaxies is chemically
homogeneous. Therefore, the new metals processed and injected by the current
star formation episode are possibly not observed and reside in the hot gas
phase, whereas the metals from previous events are well mixed and homogeneously
distributed through the whole extent of the galaxy.

\end{abstract}


\section{Introduction}

HII galaxies (or Blue Compact Dwarfs, BCDs) are low-mass and metal-poor
galaxies (1/50-1/3Z$_{\odot}$), experiencing strong episodes of star
formation, characterized by the presence of bright emission lines on a faint
blue continuum \citep{T91,K04}. Several studies \citep[e.g,][]{P96,tt97,C03,Wes04}
have shown the existence of an old stellar
population underlying the present burst in most of these galaxies,
indicating that these objects are not young systems forming their first generation
of stars.

The star formation activity in HII galaxies is concentrated in several luminous
star clusters ($\sim$1-30 pc in size) spread over the galaxies. Some of these
clusters have been associated with super star clusters (SSCs), similar to those
found in interacting galaxies, such as the antennae \citep{W95,W99}, and some
local dwarf galaxies such as Henize 2-10 \citep{V94} and NGC 1569
\citep[e.g.,][]{O94,AS85}. During their evolution, these clusters ionize the
interstellar medium (ISM) causing the formation of giant HII regions (GHIIRs),
and release a considerable quantity of mechanical energy into the ISM,
characterized by supersonic motions \citep[e.g,][]{M87}, sweeping up the
surrounding medium, removing the gas and dust from the star formation site,
while producing structures such as bubbles and supershells. Ultimately,  the
evolution of these young clusters also causes the freshly produced metals to be dumped
into the ISM.

Oxygen and other $\alpha$ elements are produced by massive stars ($>$8-10
M$_{\odot}$) and are released into the ISM during their supernova phase. These
metals will be dispersed in the whole galaxy and mixed via hydrodynamic
mechanisms in time scales of few 10$^{8}$ yrs \citep[e.g.,][]{T96}. 
Since the spatial distribution of abundances of O, N, etc depends on the
recycling time of the ISM, the spatial variation of these abundances give
important insights about these processes. The spatial distribution of emission
line ratios and of the abundance of certain elements in dwarf galaxies have been
studied by different authors \citep[e.g.,][]{K96, K97b, Lee06}. In these works
the radial and spatial distribution of the O/H abundance, obtained from the
analysis of their H II regions, have been used as a chemical evolution
indicator. Localized nitrogen self enrichment has been measured in a few cases
and attributed to the action of strong winds produced by Wolf-Rayet (W-R) stars
\citep[e.g.,][]{W89, K97a}, but no oxygen localized enrichment systems have been
confirmed \citep{K96}. In GHIIRs, where only one star formation region is
present, it is usually assumed that the abundance of the galaxy and their ISM
is well represented by the abundance of this single region, assuming that the
metals are well mixed and distributed across the galaxy. 

The details of the structure of HII galaxies are important to understand their
ionization mechanisms, star formation feedback and chemical enrichment. These
issues have been actively addressed in studies of the brightest and nearest
systems. The large and heterogeneous HII galaxy sample \citep{LT86,K88,tmt97}
includes a significant number of far less studied objects at larger distances,
with small apparent sizes and with relatively low surface brightness. Typically,
HII galaxies contain one or a few star formation regions, but H$\beta$ images
\citep{L07} of some compact objects have revealed that the central region
probably hosts a myriad of unresolved star clusters. The present facilities,
with high spatial resolution and high efficiency instruments on large telescopes
(8m), allow us to study these more distant objects to derive their observed
properties and relate them to the better known nearby galaxies, thus having a
handle on the intrinsic properties of star formation as well as possible
evolutionary effects.

\begin{figure*}[!htb]
\centering
\includegraphics[width=0.6\columnwidth]{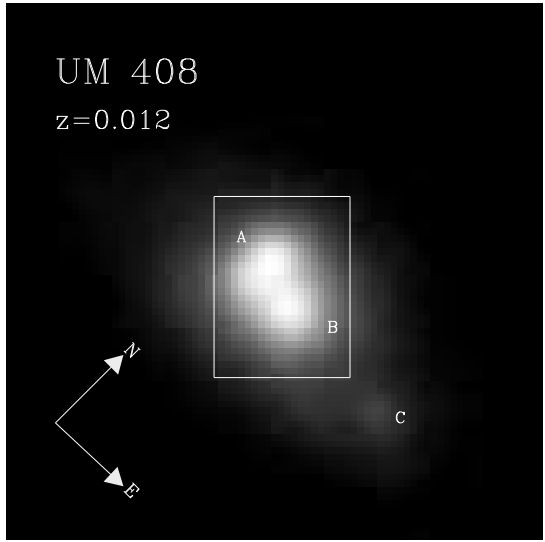}
\caption{g-band acquisition image of the galaxy UM 408. The  image is
11\arcsec.5$\times$11\arcsec.5. The rectangle indicates the field of
view of 3\arcsec$\times$4\arcsec.4  ($\sim$750 $\times$ 1100 pc) used in this 
work to construct the monochromatic maps. The two most prominent regions (A and B)
are  indicated in the image. The faint region (C in the plot)
located $\sim$3\arcsec.0 to the East from B is placed outside the field of view
(1\arcsec $\simeq$ 250 pc).} 
\label{image}
\end{figure*}

In this paper we use  integral field spectroscopy with Gemini Multi-object
Spectrograph and the Integral Field Unit (GMOS-IFU) at Gemini South in order to
study the spatial distribution of emission lines, their ratios, extinction,
abundances and the kinematic properties of the gas in the interstellar medium of
the compact dwarf galaxy UM 408. 
UM 408 belongs to a subset sample of  rare HII
galaxies with low metal abundances (i.e. Z$<$ 1/20 Z$_{\odot}$) with $12+\log
(O/H)=7.66$ \citep{M94,T95}, compact morphology,
and very small effective radius (R$_{eff}$) of only 2\arcsec.1 from previous morphology  
and surface brightness studies \citep{tt97}. On other hand, \citet{P02} using 
high resolution spectroscopy derived an abundance of $12+\log(O/H)=7.93$ and
HI observations \citep{S02} reveal
an HI mass of log(M$_{HI}$)=8.815 M$_{\odot}$. UM 408 is cataloged by \cite{S89}
as a Dwarf HII Hot Spot with a redshift of 0.012\footnote[1]{Value obtained from
NED} (v=3598 \kms) with coordinates RA$=02^{h}$ 11$^{m}$ 23\fs4, DEC$=02^{o}$
20$^{'}$ 30\arcsec (J2000) at a distance of $\sim$ 46 Mpc, and V apparent
magnitude of 17.38. With the present observations,
we note that the central part of UM 408 encompasses two main star formation
regions (here named A and B), as shown by the g-band acquisition images in
Figure \ref{image}. Another faint region, called C, can be seen in the outer
parts of the galaxy. These regions were all unresolved in previous studies.

The observations and data reduction procedures  are discussed in Section
\ref{obser}. Section \ref{results} gives the results obtained from the 2D
emission line maps. In Section \ref{discussion} we discuss our results and our
conclusions are presented in section \ref{conclusions}. \section{Observations
and Data Reductions}\label{obser}

The observations were performed with the Gemini Multi-Object Spectrograph GMOS
\citep{Hook04} and the IFU unit \citep[][hereafter GMOS--IFU]{Allington02} at
Gemini South Telescope in Chile during the night of August 21 and December 31
2004, using the grating B600$+\_$G5323 (B600) and on January 01, 2005 with the
grating R600$+\_$G5324 (R600) in one slit mode, covering a total spectral range
from $\sim$3021 to 7225 $\rm\AA$. The GMOS-IFU in one slit mode composes a
pattern of 750 hexagonal elements, each with a projected diameter of 0\arcsec.2,
covering a total 3\arcsec.5 $\times$5$\arcsec$ field of view, where 250 of these
elements are dedicated to sky observation. The detector is formed by three
2048$\times$4608 CCDs with 13.5 $\mu$m pixels, with a scale of
0.073\arcsec/pixel. The CCDs create a mosaic of 6144$\times$4608
pixels with a small gap between the chips of 37 columns.

\begin{figure*}[!htb]
\centering
\includegraphics*[width=0.49\columnwidth]{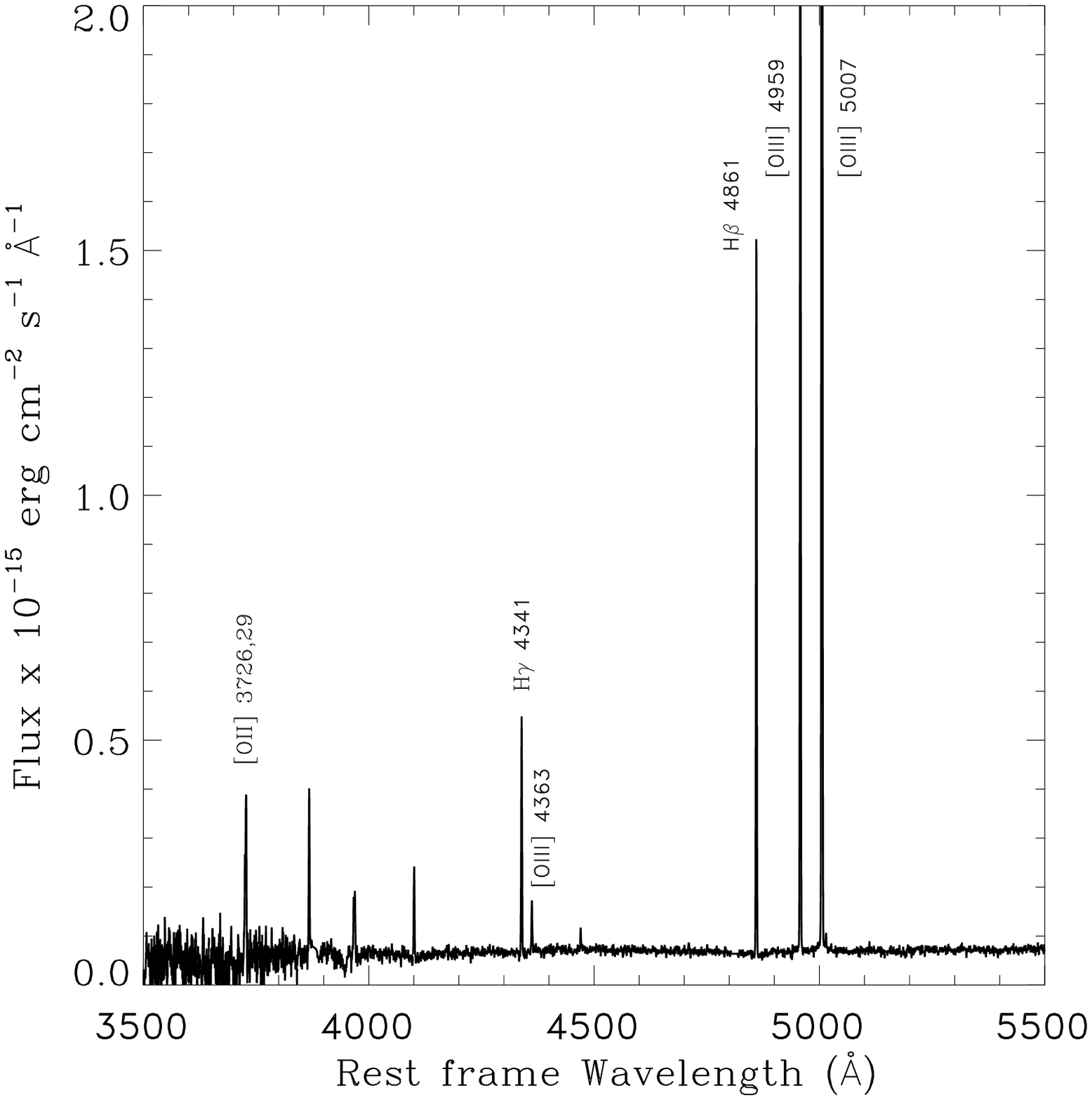}
\includegraphics*[width=0.49\columnwidth]{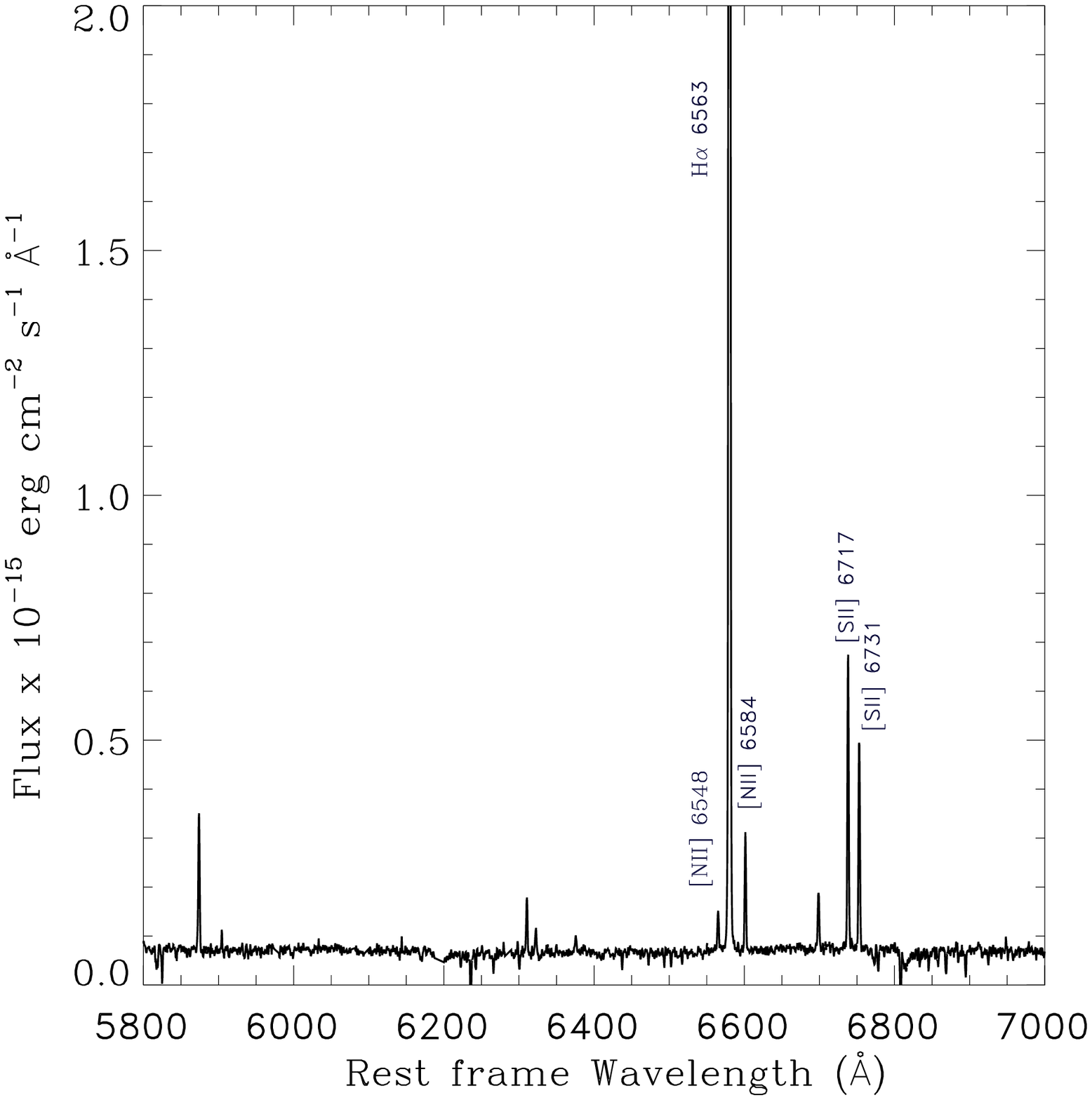}
\caption{1D blue and red parts of the spectrum, obtained as the sum of all fibers, in 
the rest-frame wavelength. In this Figure, only the lines used in the analysis are indicated.
} \label{spec}
\end{figure*}

\begin{figure*}[!htb]
\centering
\includegraphics*[width=0.8\columnwidth]{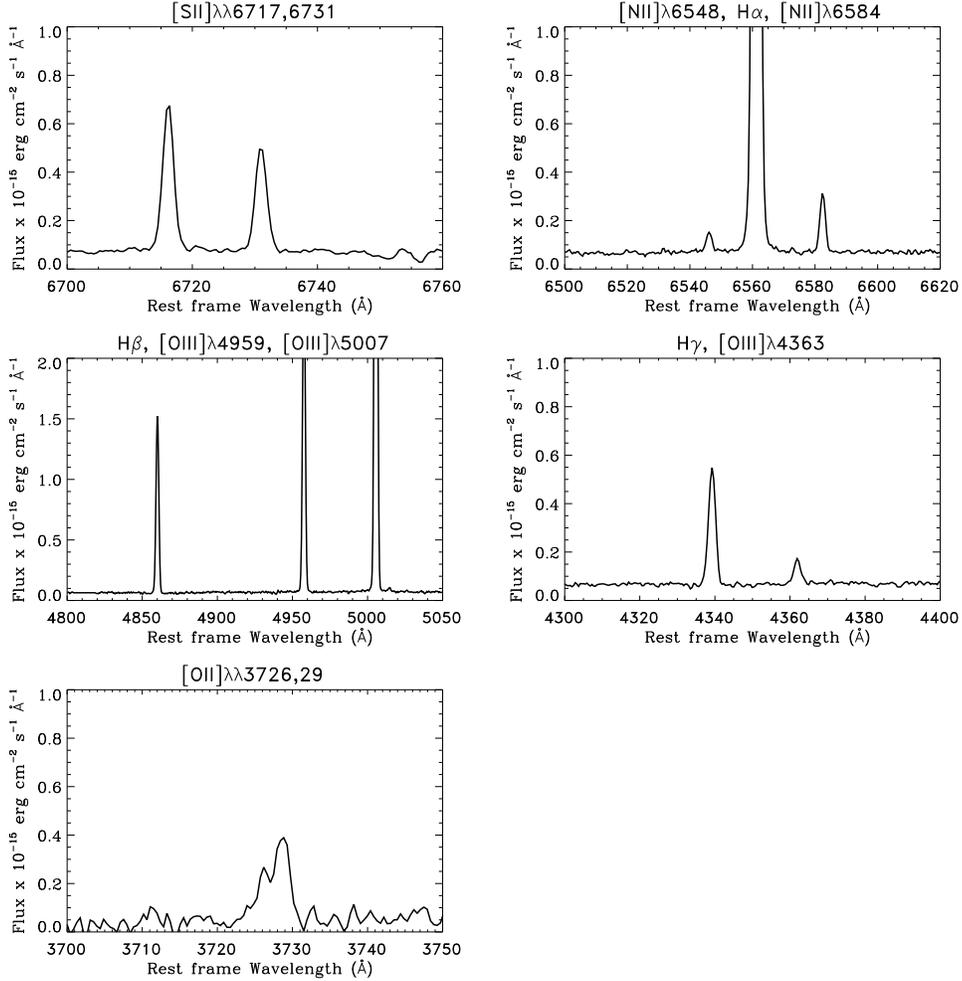}
\caption{Subsets from the spectra in Figure 2 showing details of selected emission lines considered in
this study.
} \label{speczoom}
\end{figure*}

Table 1 shows the  observing log, indicating the instrumental setup, airmass and
exposure times, the dispersion, the final resolution and the seeing (FWHM) of
each observation. The data were reduced using the Gemini package version 1.8
inside IRAF\footnote[2]{IRAF is distributed  by NOAO, which is operated by the
Association of Universities for Research in Astronomy Inc., under cooperative
agreement with the National Science Foundation.}. All science exposures,
comparison lamps and spectroscopic twilight and GCAL flats were overscan/bias
subtracted and trimmed. The spectroscopic GCAL flats were processed by removing
the calibration unit plus GMOS spectral response and the calibration unit uneven
illumination. Twilight flats were used to correct for the illumination pattern
in the GCAL lamp flat using the task gfresponse in the GMOS package. The
twilight spectra were divided by the response map obtained from the lamp flats
and the resulting spectra were averaged in the dispersion direction, giving the 
ratio of sky to lamp  response for each fiber. The final response maps were
obtained then by multiplying  the GCAL lamp flat by the derived ratio. The
resulting extracted spectra were then wavelength calibrated, corrected by the
relative fiber throughputs, and extracted. The residual values in the wavelength
solution for 60-70 points using a 4$^{th}$ or 5$^{th}$  Chebyshev polynomial
typically yielded \textit{rms} values of $\sim$0.10~\AA\ and $\sim$0.15~\AA\ for
both gratings respectively. The final spectra cover a wavelength interval of
$\sim$3021--5823 \AA\ and $\sim$4371--7225 \AA\ for data taken with the B600 and
R600 gratings, respectively. The final spectral resolutions and dispersions are
shown in Table 1 (columns 5 and 6).

The flux calibration was performed using the sensitive function derived from
observation of the star LTT 3864 for both gratings in December 31 and January 1.
Standard stars were not observed in August. Therefore, the blue part of the
spectra was constructed using only the observation obtained in December 2004.
The 2D data images were transformed into 3D data cube (x,y,$\lambda$) using the
\textit{gfcube} routine and resampled as square pixels with 0\arcsec.1  of
spatial resolution. The emission line fluxes were measured using the IRAF task
\textit{fitprofs} by fitting Gaussian profiles.  From these, we then created the
emission line maps used in our analysis (see Section~\ref{results}).

At shorter wavelength the IFU spectroscopy and long-slit observations suffer a
spatial translation produced by differential atmospheric refraction (DAR). This
effect is wavelength dependent, and is produced by the deviation of the light
when it passes through the atmosphere due to the variation of the air density as
a function of elevation. 
In order to obtain fluxes corrected of DAR, we used a similar procedure
as described in \citet{A99} and also applied by \citet{W07a}. First we splitted
the data cube in monochromatic images, and for each image we calculated the centroid of 
an unresolved point source in the field of view, that in our case corresponds to region A. 
Finally, the various monochromatic maps were aligned by shifting the centroids to the same position.

In Figure \ref{spec} we show the resultant spectrum for the two gratings used, obtained
from the sum of all GMOS-IFU fibers. Figure \ref{speczoom} shows a detailed view
of the selected emission lines considered in this study. The spectral resolution
of the GMOS-IFU observations allow us to resolve the [OII]$\lambda\lambda$3726,29
doublet lines in a limited number of apertures with a $\Delta\lambda$=2.32 $\rm
\AA$ between the peak of the lines in the integrated spectrum (see Figure
\ref{speczoom}). The low intensity of [OII]$\lambda\lambda$3726,29 lines and
[OIII]$\lambda$4363 are the largest sources of uncertainties in our
results. An automatic procedure to measure these emission line fluxes was not
possible, therefore these lines were measured manually using \textit{splot}.

\section{Results}\label{results}

\subsection{Emission line, continuum and EW(H$\beta$) maps}

\begin{figure*}
\centering
\includegraphics*[width=0.27\columnwidth]{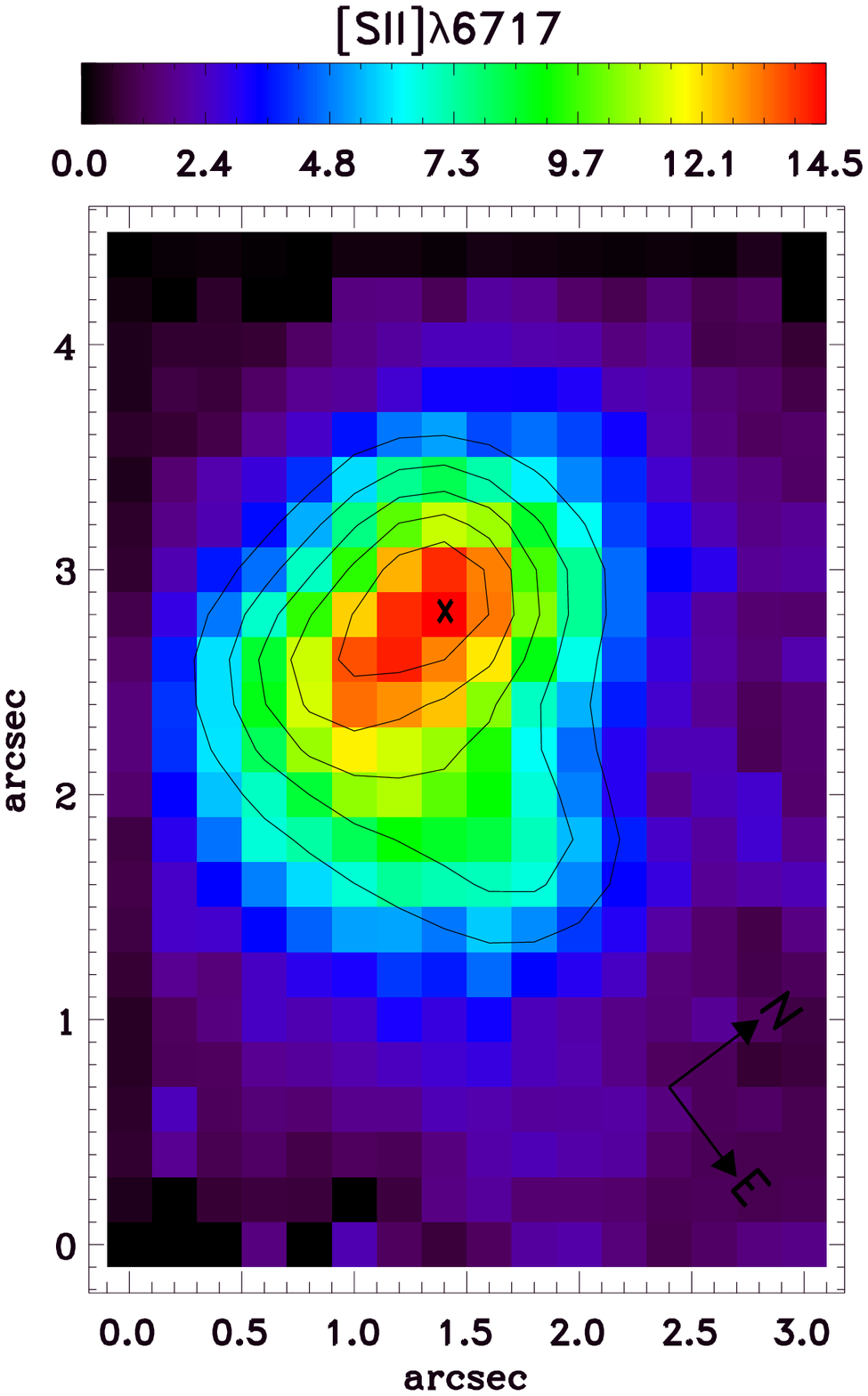}
\includegraphics*[width=0.27\columnwidth]{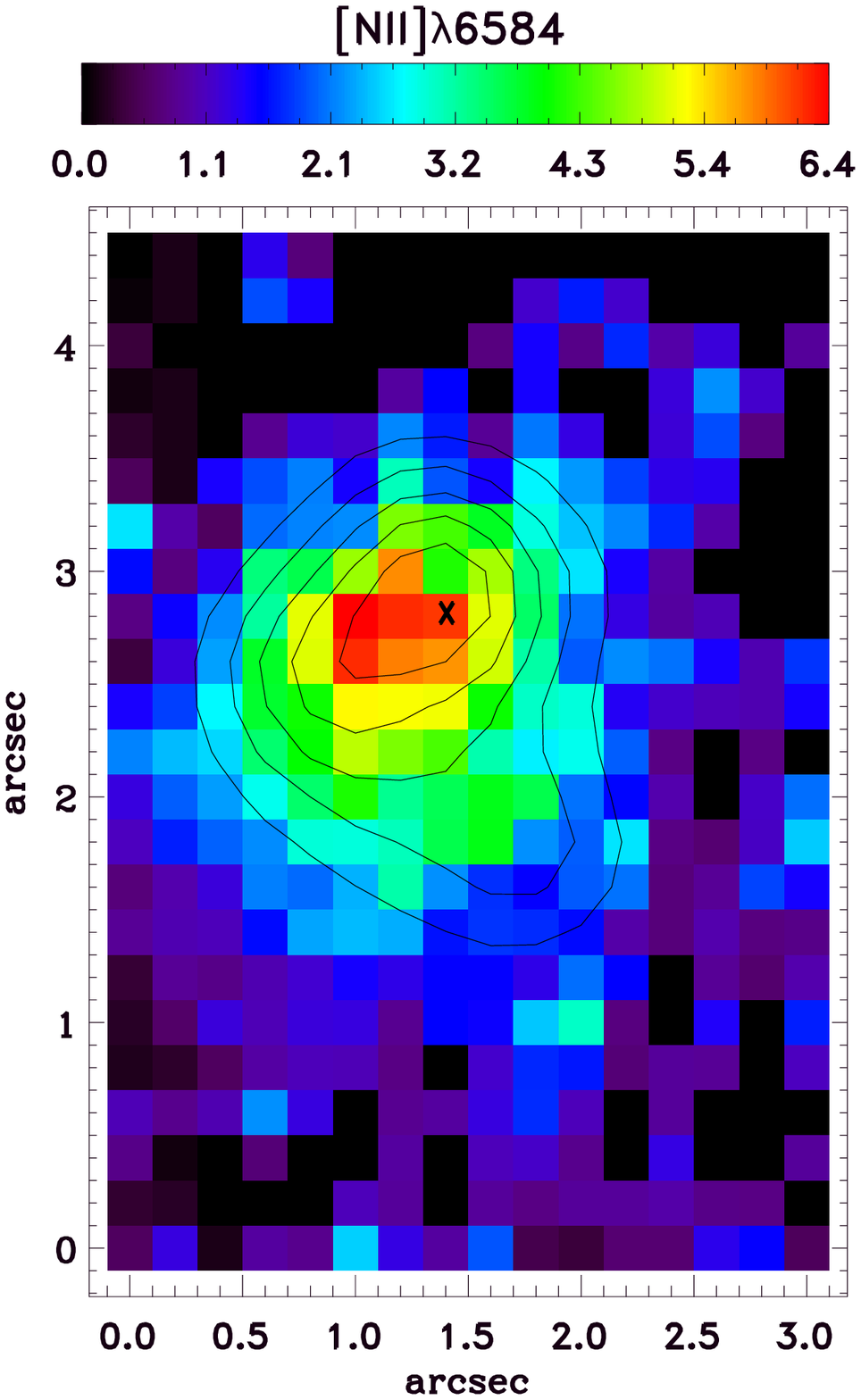}
\includegraphics*[width=0.27\columnwidth]{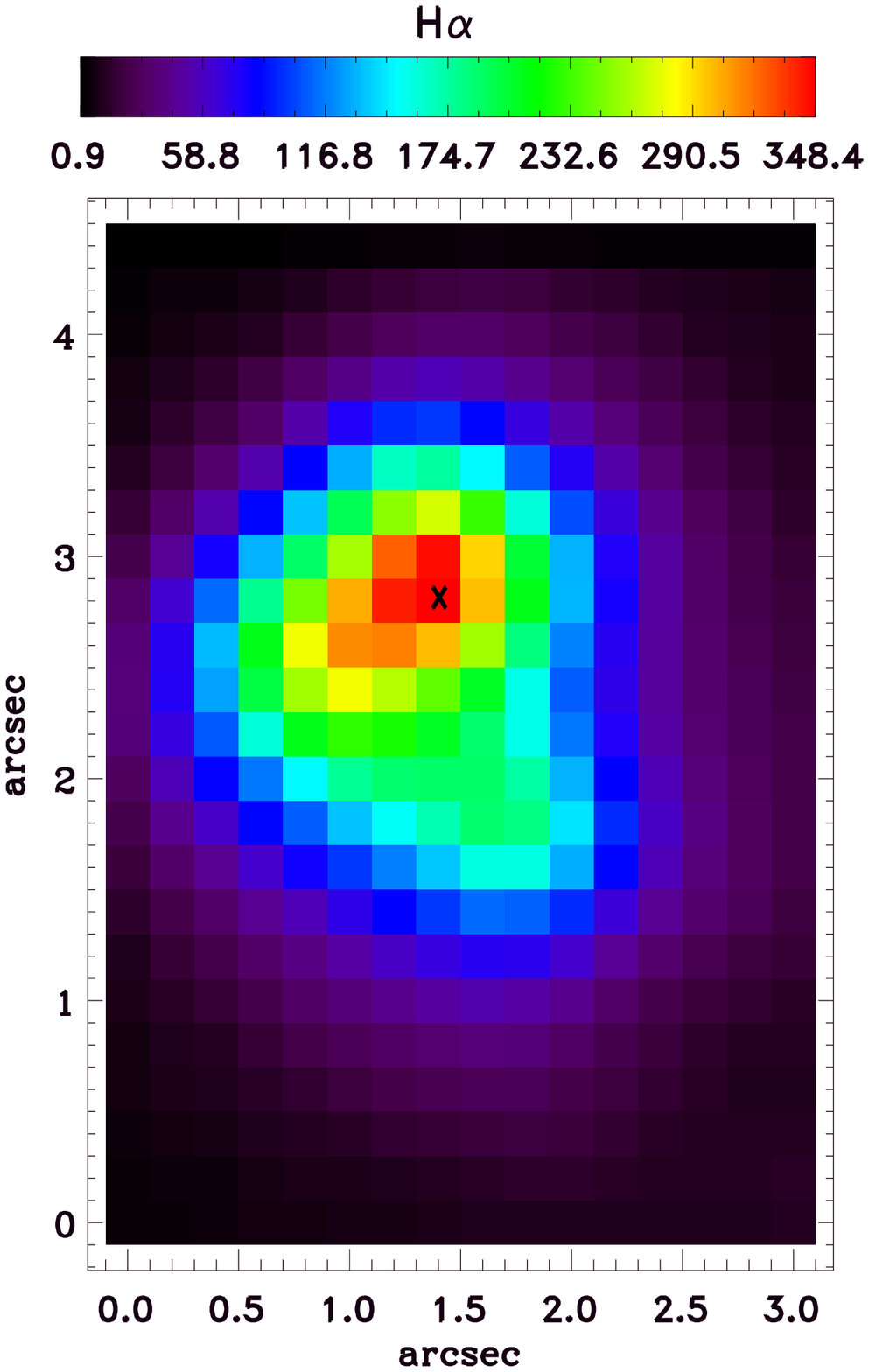}

\includegraphics*[width=0.27\columnwidth]{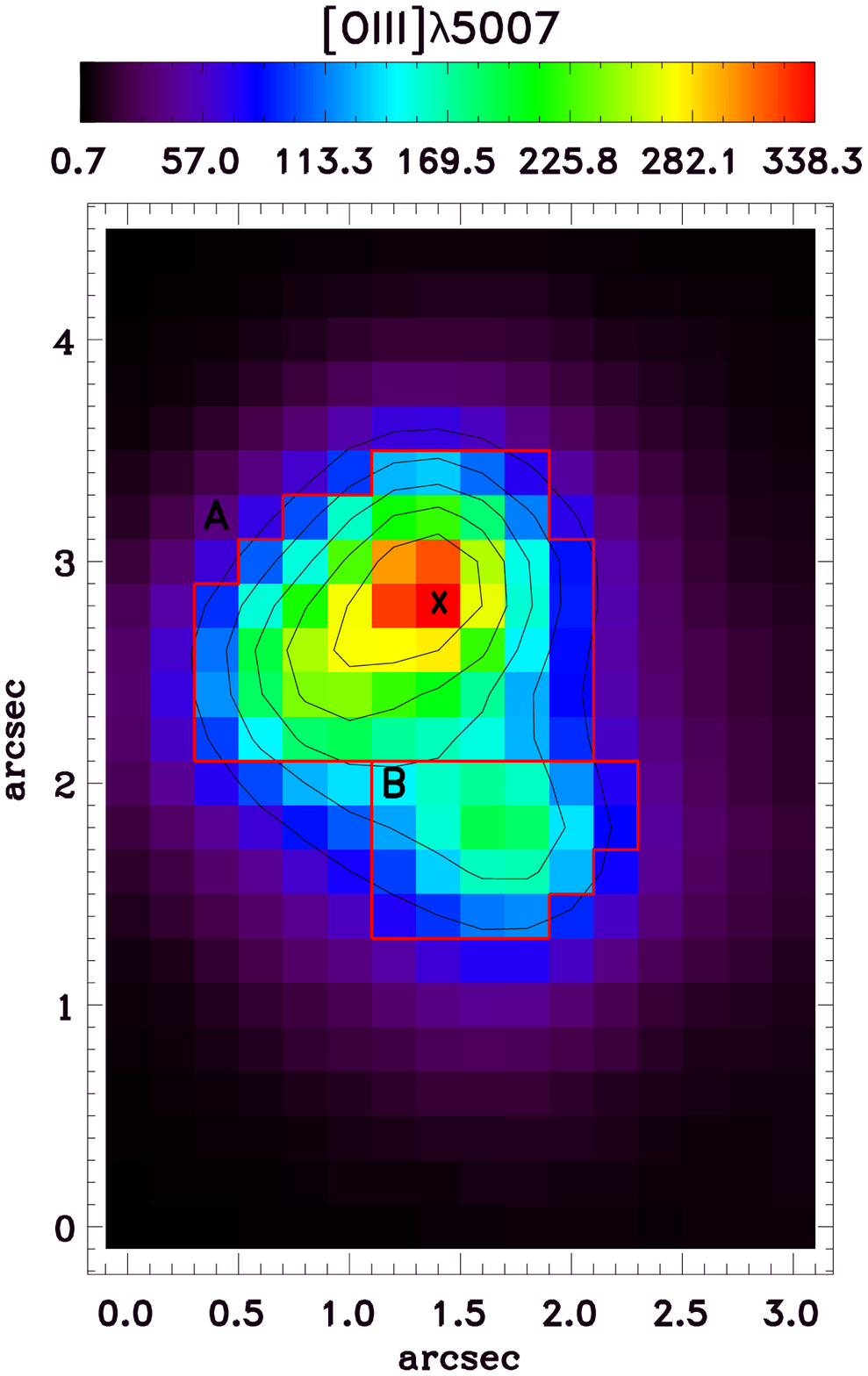}
\includegraphics*[width=0.27\columnwidth]{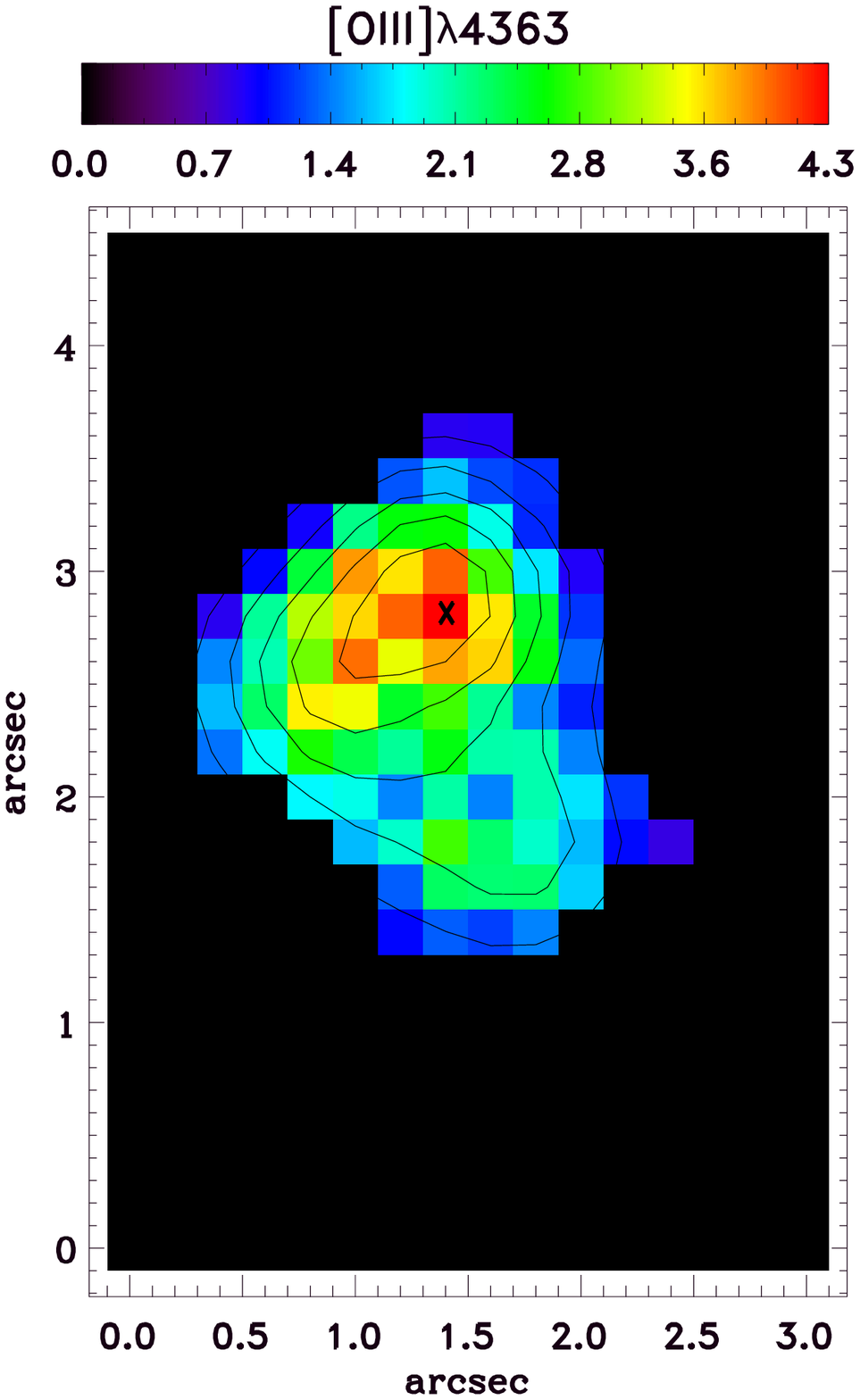}
\includegraphics*[width=0.27\columnwidth]{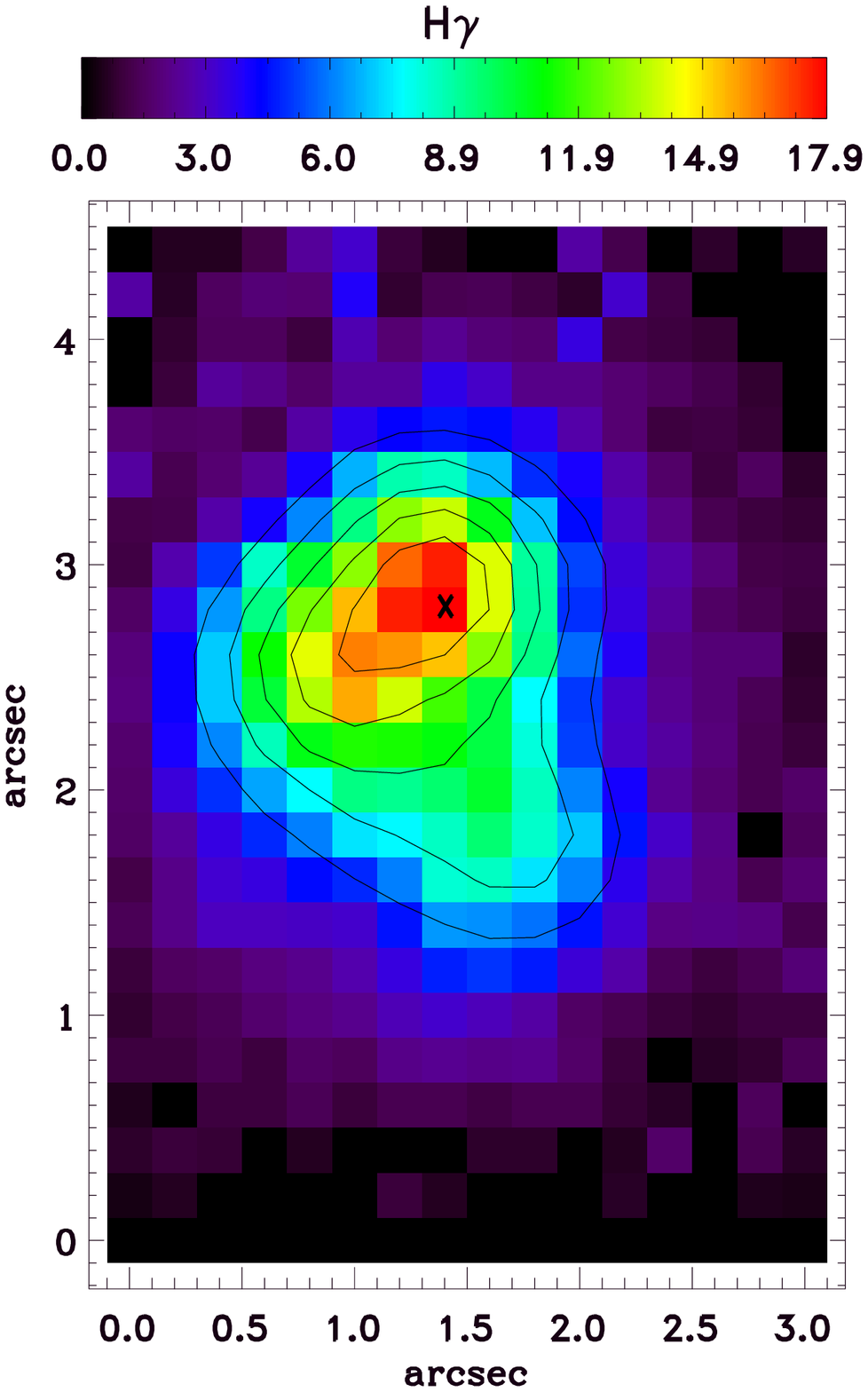}

\includegraphics*[width=0.27\columnwidth]{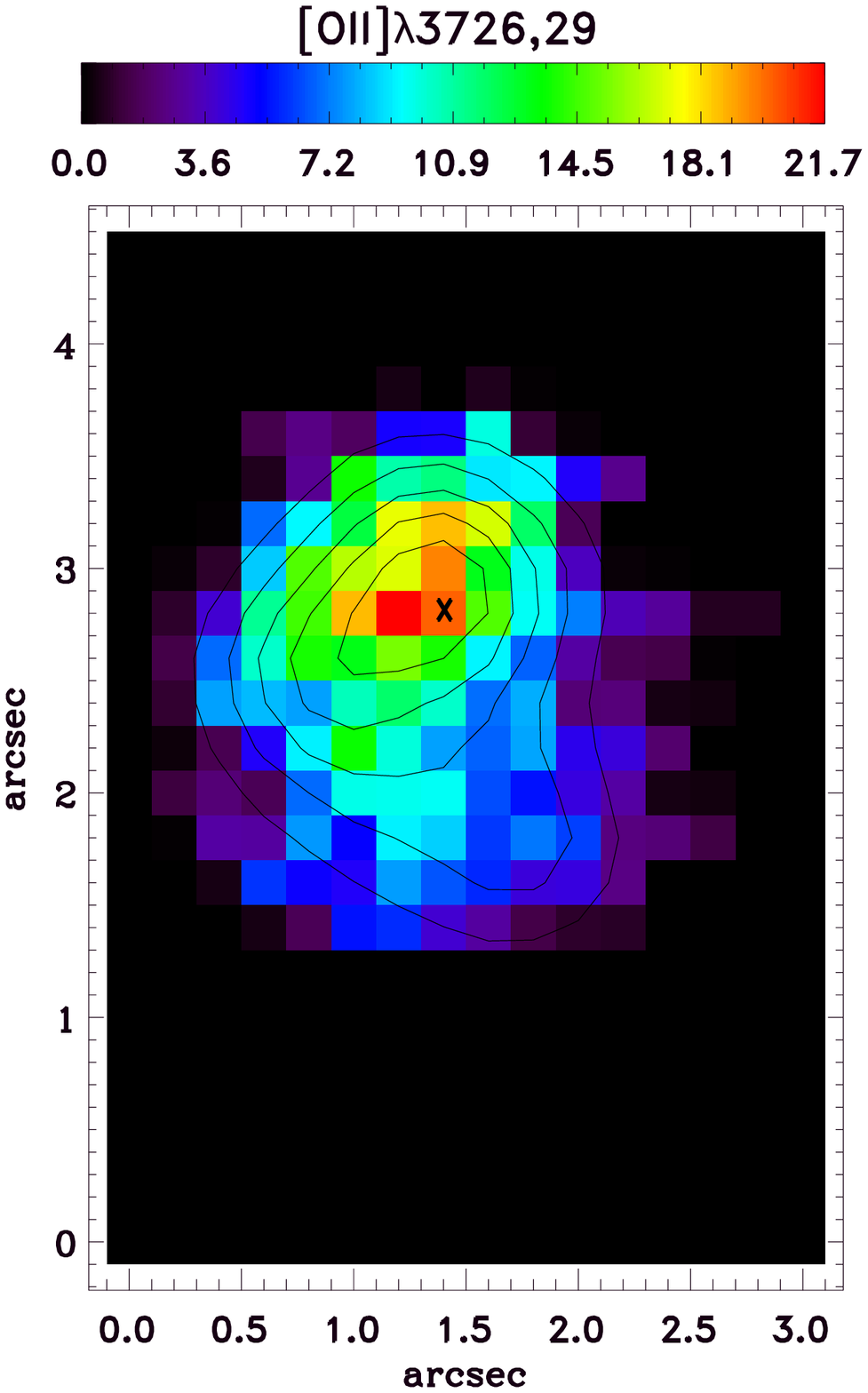}
\includegraphics*[width=0.27\columnwidth]{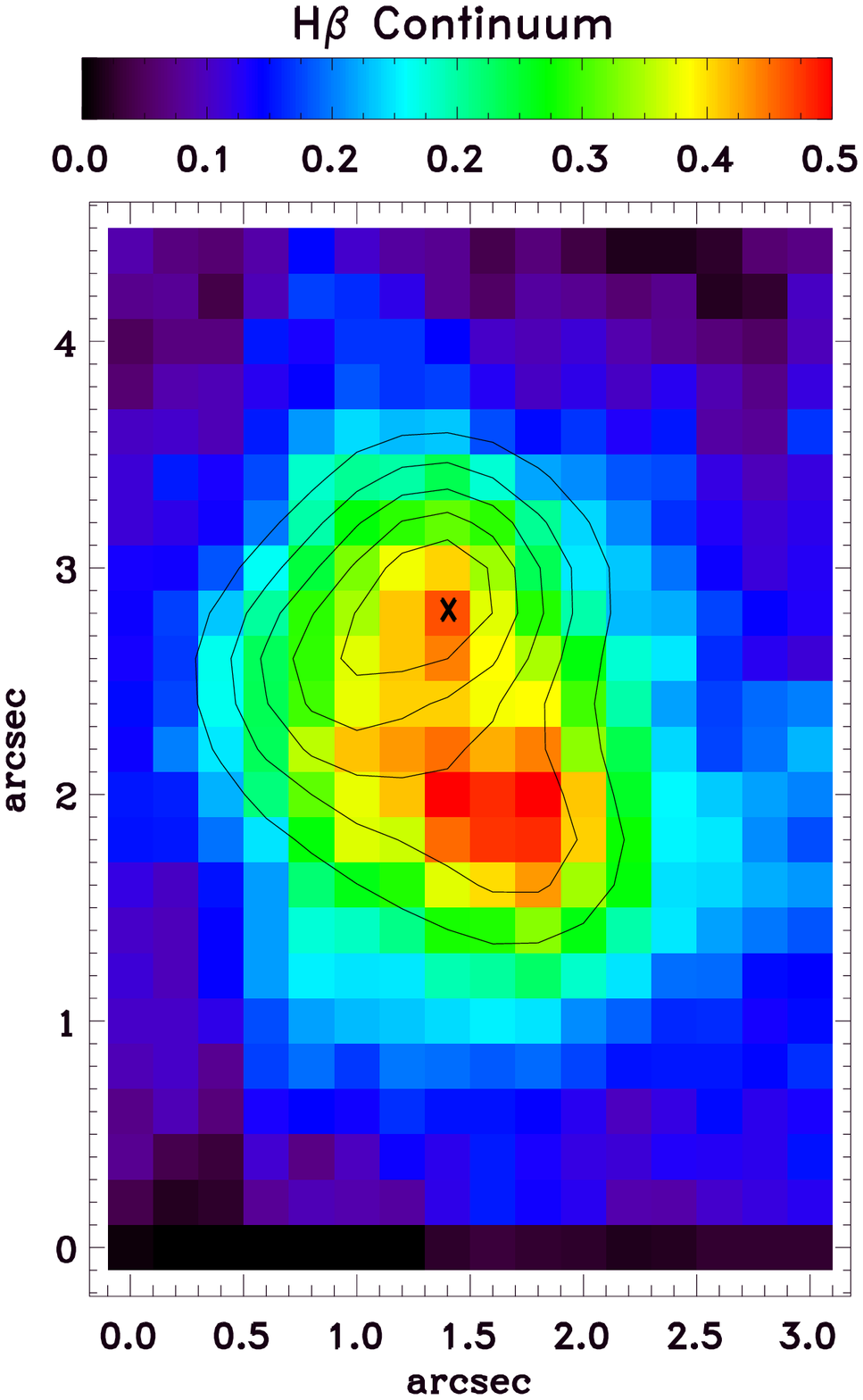}
\includegraphics*[width=0.27\columnwidth]{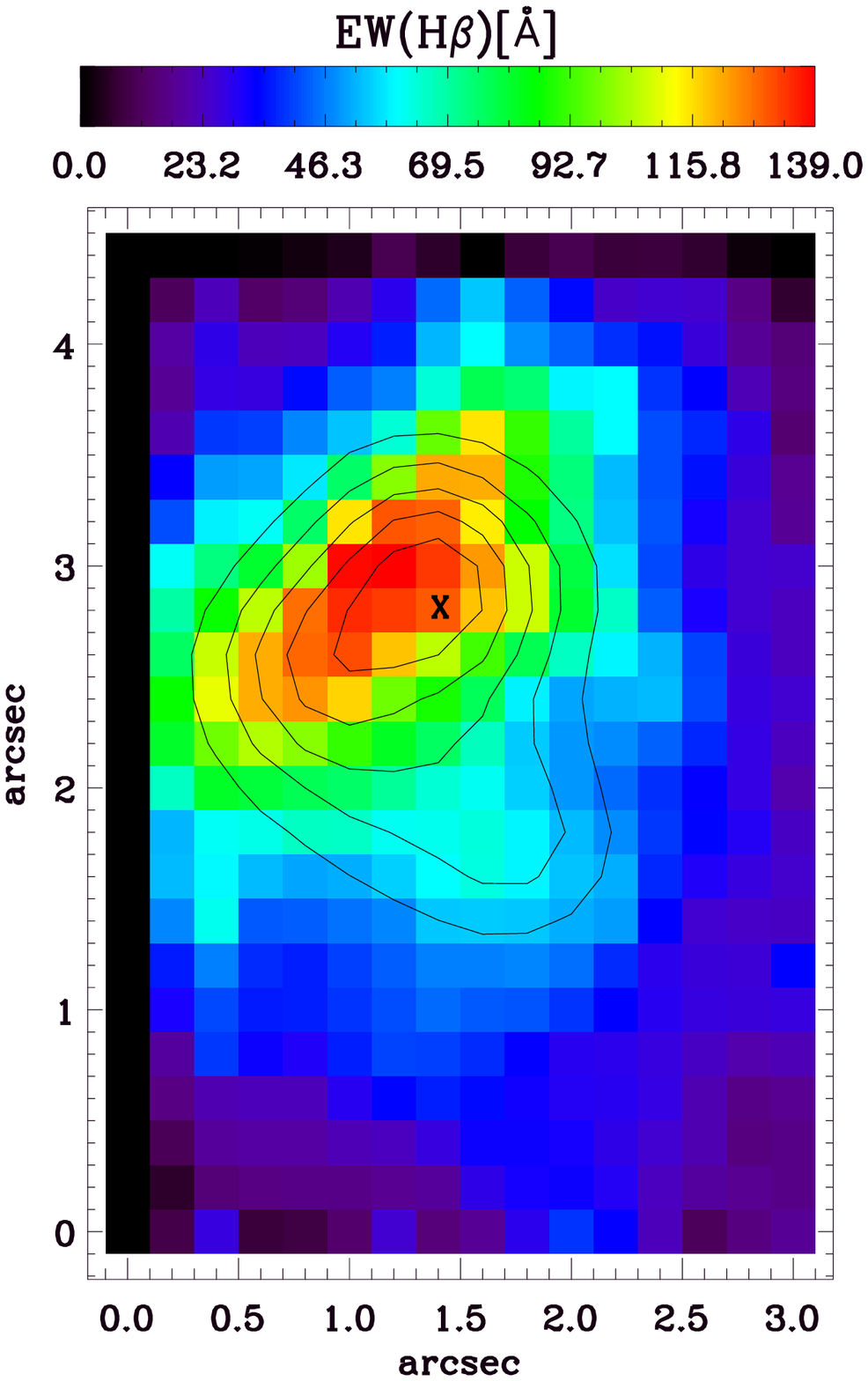}

\caption{Observed maps for
[SII]$\lambda$6717, [NII]$\lambda$6584, H$\alpha$, [OIII]$\lambda$5007,
[OIII]$\lambda$4363, H$\gamma$ and [OII]$\lambda\lambda$3726,29 emission lines, H$\beta$ continuum and
EW(H$\beta$). Fluxes in units of  $\times$10$^{-18}$ erg
cm$^{-2}$ s$^{-1}$ and EW(H$\beta$) in units of $\rm \AA$. H$\alpha$ emission line contours are overplotted on all maps. The maximum H$\alpha$ emission is placed over region A and is indicated in the maps by an X symbol. The regions in the [OIII]$\lambda$5007 map mark the areas considered as regions A and B.}
\label{line1}
\end{figure*}

We used a flux measurement procedure described in the previous section to
produce the maps for the following emission lines: [SII]$\lambda$6731,
[SII]$\lambda$6717, [NII]$\lambda$6584, H$\alpha$, [OIII]$\lambda$5007,
[OIII]$\lambda$4959, H$\beta$, [OIII]$\lambda$4363, H$\gamma$,
[OII]$\lambda\lambda$3726,29, H$\alpha$ and H$\beta$ continuum and EW(H$\beta$), in a
rectangular field of 3\arcsec$\times$4$\arcsec$.4 (see Figure \ref{image}). In
order to reduce the number of data points without loosing spatial information
and S/N, the data were resampled to 0$\arcsec$.2 in spatial resolution. Typical
S/N ratios for individual pixels (spaxels in our data cube) for
[OIII]$\lambda$5007 are $\gtrsim$700 and $\gtrsim$ 500 for regions A and B,
respectively. The [OIII]$\lambda$4363 line shows a typical S/N ratio $\gtrsim$8
and $\gtrsim$4 for regions A and B, respectively. The signal to noise for each
emission line is given by the ratio between the emission line peak and the RMS
of the adjacent continuum. Figure \ref{line1} shows some of the observed
emission line maps. All emission and continuum maps in this Figure are in
units of $\times$ 10$^{-18}$ erg cm$^{-2}$ s$^{-1}$, except for the H$\beta$
equivalent width map, which is in units of $\rm \AA$. Here 0$\arcsec$.2 $\approx$ 50 pc
(distances in this work were computed assuming a Hubble constant of H$_{0}$=72 km s$^{-1}$ Mpc$^{-1}$).
Since most of the emission lines were measured using an automatic procedure, we 
filtered the maps assigning the value 0 erg cm$^{-2}$s$^{-1}$ to all pixels with S/N$<$3.

The maps in Figure \ref{line1} show essentially the same spatial distribution,
presenting two well defined and regular regions, that we have labeled A and B. The maximum
peak intensity of the H$\alpha$ emission line in region A is indicated in the
maps by an X symbol. The recombination lines H$\alpha$, H$\beta$ and H$\gamma$
have very similar spatial distribution to the forbidden lines in Figure
\ref{line1}, but there is a slight  difference in the spatial distribution of
[OIII]$\lambda$4363 and [OII]$\lambda\lambda$3726,29 over region B. The last two images in Figure
\ref{line1}  corresponds to the spatial distribution of the H$\beta$ continuum
and the EW(H$\beta$), respectively. It is clear from the maps that the continuum
peak is not coincident with the maximum in H$\alpha$ emission. This has been
previously reported by other authors \citep[e.g.,][]{S91,M98,L07} in some
galaxies with single and multiple star forming regions. In NGC 4214,
\citet{S91} and \citet{M98} found that the continuum peak emission was
displaced in relation to the line emitting regions where a strong Wolf-Rayet (WR) feature
(HeII $\lambda$4686~\AA) was also seen. No WR  signatures were 
detected in the spectra of UM 408.

The largest EW(H$\beta$) values  in the galaxy are associated with  region A,
indicating that this region is younger than region B. On the other hand, the
H$\beta$ continuum map (see Figure \ref{line1}) shows the largest values over
region B. In Section~\ref{discussion}, the EW(H$\beta$) values are used as a proxy
for the starburst ages  in order to derive
the integrated physical properties of the galaxy and their star forming regions. 
The error associated to each emission line ($\sigma_{i}$) was calculated using
the relationship given in \citet{C02}:

\begin{equation}
\sigma_{i}=\sigma_{c} N^{1/2} \sqrt{1+\frac{EW}{N\bigtriangleup}},
\end{equation}

\noindent
where $\sigma_{c}$ is  the standard deviation in the local continuum
associated to each emission line, $N$ is the number of pixels, EW is the
equivalent width of the line and $\bigtriangleup$ is the instrumental
dispersion in $\rm \AA$ (see Table 1 \label{obslog}).  These estimated errors will
be used later in the derivation of uncertainties in the oxygen abundance 
determination and integrated properties.

Finally, given that the seeing was 0.8-0.9$\arcsec$ ($\sim$4 pixels in the emission line maps), each aperture or pixel of 0.2$\arcsec$ should be correlated with its neighbor, and the pixel-to-pixel variations in the maps produced from the observed emission line are likely due to the uncertainties attached to the measurement and do not represent real variations. Therefore, we smoothed the emission line ratio, temperature and oxygen abundance maps using a 3$\times$3 box.

\subsection{Extinction}

\begin{figure}[!hbt]
\centering
\includegraphics*[width=0.5\columnwidth]{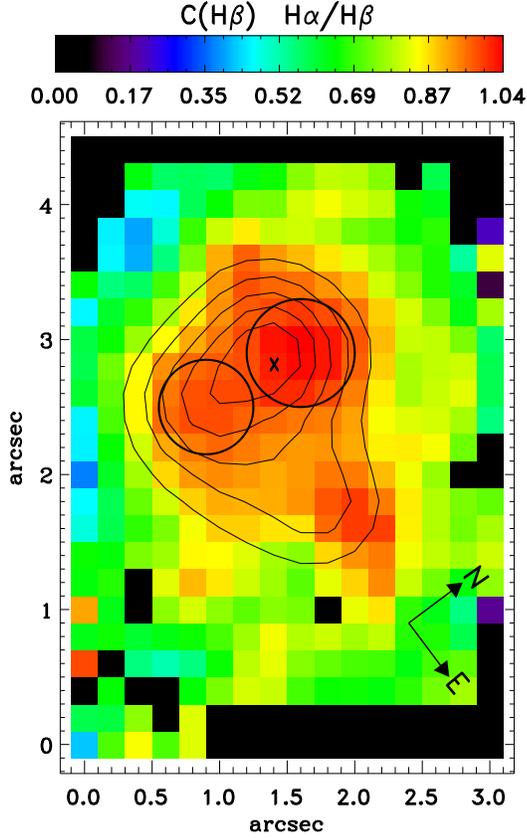}
\caption{Extinction coefficient distribution, obtained from the emission lines H$\alpha$/H$\beta$ ratios. The regions enclosed by circles mark the location of other extinction peaks, probably related to other unresolved star clusters. The maximum
H$\alpha$ emission is placed over region A and is indicated by an X symbol. H$\alpha$ contours are overplotted on the maps.}
\label{extinction}
\end{figure}

The logarithmic reddening parameter c(H$\beta$) was calculated from the ratio
H$\alpha$/H$\beta$, where the intrinsic
value of 2.87 for case B recombination was assumed. 
Thus, the corrected emission lines are calculated as

\begin{equation}\label{c(hb)}
\frac{I(\lambda)}{I(H\beta)}=\frac{F(\lambda)}{F(H\beta)} \times 10^{c(H\beta)f(\lambda)},
\end{equation}\noindent

\noindent
where I($\lambda$) and F($\lambda$) are the dereddened flux and observed flux
at a given wavelength, respectively and f($\lambda$) is the reddening
function given by \citet{S79} as parametrized by \citet{H83}, using appropriated values
for  the Milky Way \citep{K97a}.

Figure \ref{extinction} shows the smoothed spatial distribution of c(H$\beta$)
for the emission line ratios mentioned above. The reddening parameters
c(H$\beta$) were set to 0.0 for unrealistic values of H$\alpha$/H$\beta$ $<$
2.87. We observe a range of extinction of $\sim$0.02 to 1.04 with average
c(H$\beta$) value of 0.76 and a standard deviation of 0.17. The lowest
extinction values are located in the outer part of the field of view (see Figure
\ref{extinction}). We note also that the maximum extinction c(H$\beta$) is close
to, but not coincident with the peak of H$\alpha$ emission. A similar result has
been reported by \citet{M98} and \citet{CA02} for the star formation knots in
the galaxies NGC 4214 and Mrk 370, respectively. Apparently, the current
starburst episode in this galaxy is sweeping the gas and dust out of the center
into the surrounding regions. The emission line distribution as well as the
c(H$\beta$) distribution over region A and its elongated shape suggest that this
region may contain two or more star forming complexes instead of only one. These
complexes are indicated with circles in the extinction maps.

Finally, as the extinction across the galaxy does not show a uniform distribution, 
concentrating the highest values around the star forming regions, the fluxes of 
each pixel in the emission line maps were corrected using their corresponding c(H$\beta$) value.

\subsection{Nebular emission line diagnostic diagrams}

\begin{figure*}[!h]
\centering
\includegraphics[width=5cm]{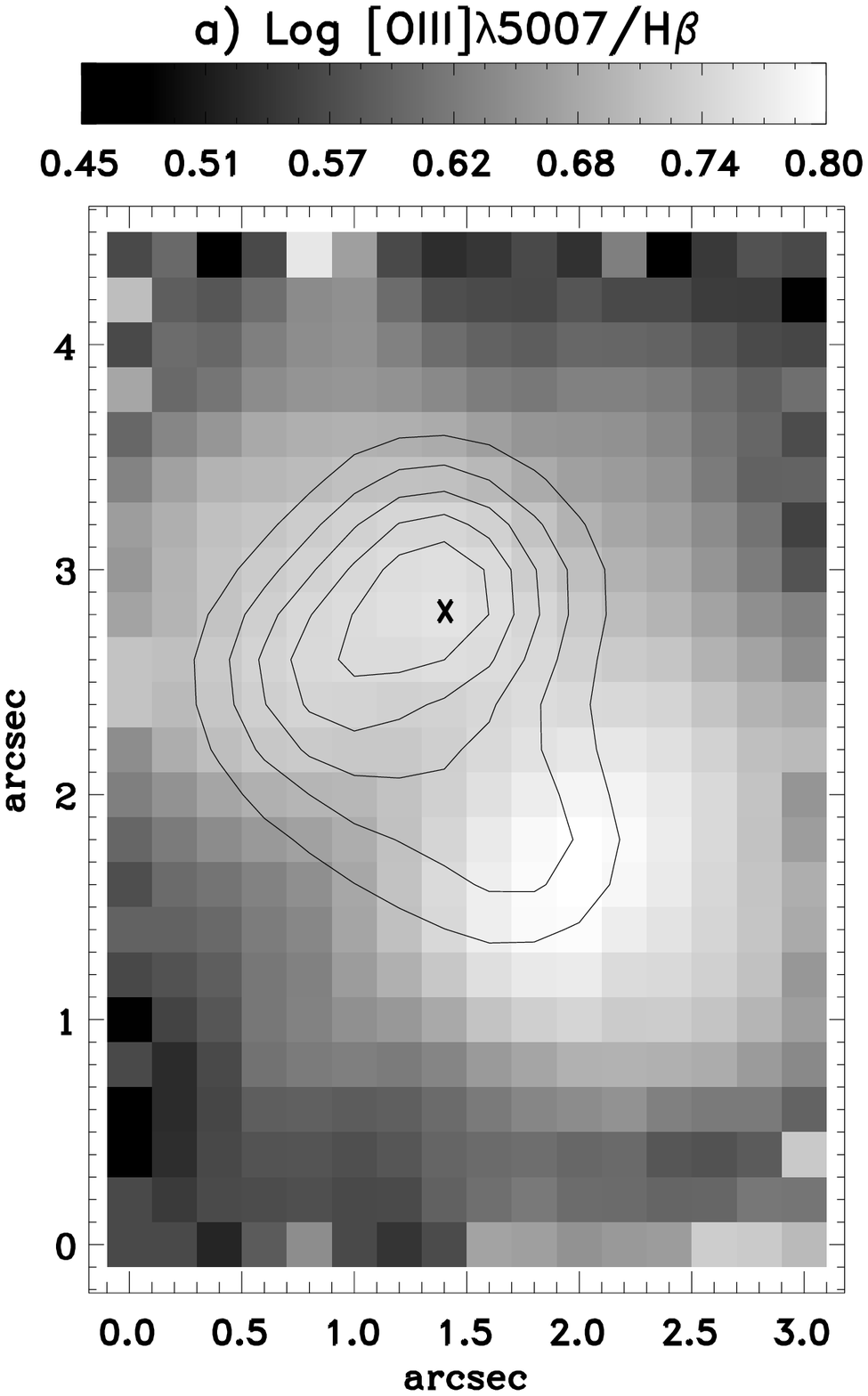}
\includegraphics[width=5cm]{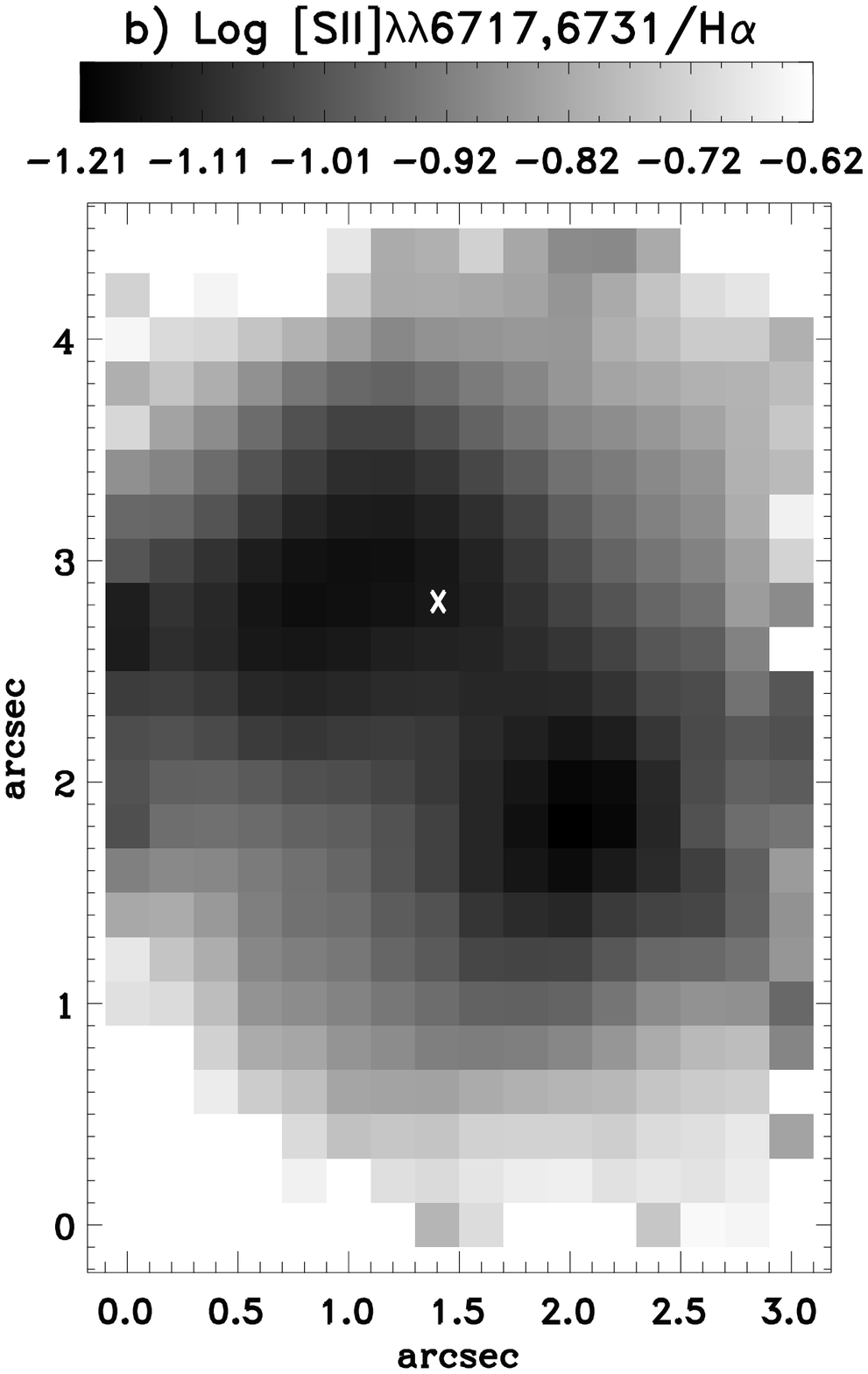}
\includegraphics[width=5cm]{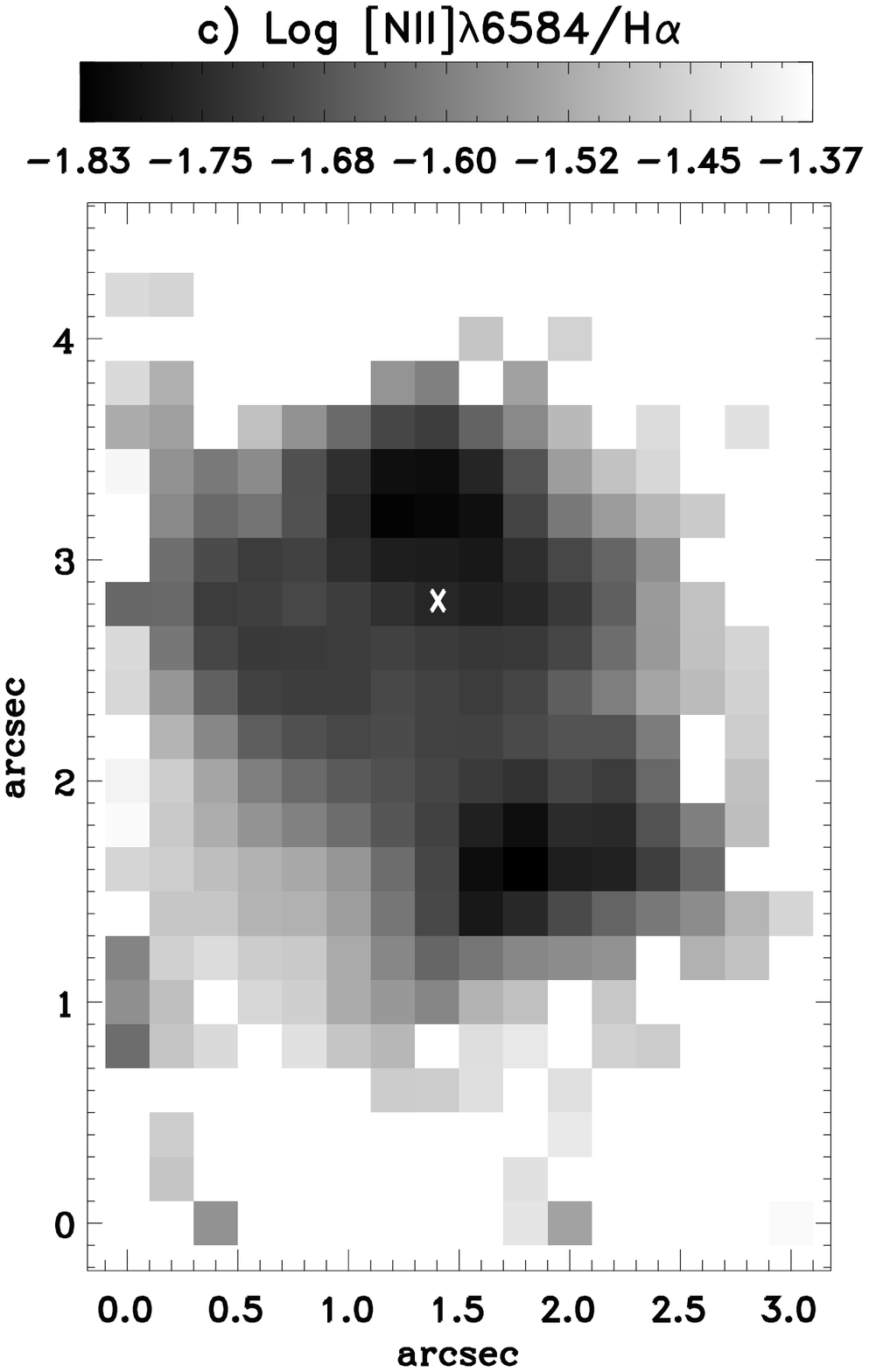}
\caption{Emission line ratios: (a) Log [OIII]$\lambda$5007/H$\beta$, 
(b) Log [SII]$\lambda\lambda$6717,6731/H$\alpha$ and (c) Log [NII]$\lambda$6584/H$\alpha$. The maximum
H$\alpha$ emission is placed over region A and is indicated in the maps by an X symbol. H$\alpha$ contours are overplotted on Figure a).}
\label{ratio}
\end{figure*}

The standard BPT diagnostic diagrams \citep{BPT} have been used to analyze the
possible excitation mechanisms present in UM 408. Figure \ref{ratio} (a, b and
c) shows the following emission line ratio maps:
[OIII]$\lambda$5007/H$\beta$ ([OIII]/H$\beta$),
[SII]$\lambda\lambda$6717,6731/H$\alpha$ ([SII]/H$\alpha$) and
[NII]$\lambda$6584/H$\alpha$ ([NII]/H$\alpha$). Figure \ref{ratio} shows that
[OIII]/H$\beta$ and [SII]/H$\alpha$ ratios have essentially the same spatial
distribution with inverse trends. The largest values of [OIII]/H$\beta$ and the
lowest values of [SII]/H$\alpha$ are seen in the east part of the galaxy, near
region B. The emission line ratios measured within these regions are common to
high excitation HII regions with values ranging from
log([OIII]/H$\beta$)=0.45 to 0.80 (all map),  0.65 to 0.76 (region A) and 0.72 to 0.80 (region B), and
from log([SII]/H$\alpha$)=-1.21 to -0.62 (all map), -1.17 to -0.85 (region A) and -1.21 to -1.01 (region
B). Additionally, the low ionization ratios [NII]/H$\alpha$ and 
[SII]/H$\alpha$ increase from the center outwards.

\begin{figure}[!htb]
\centering
\includegraphics[width=8cm]{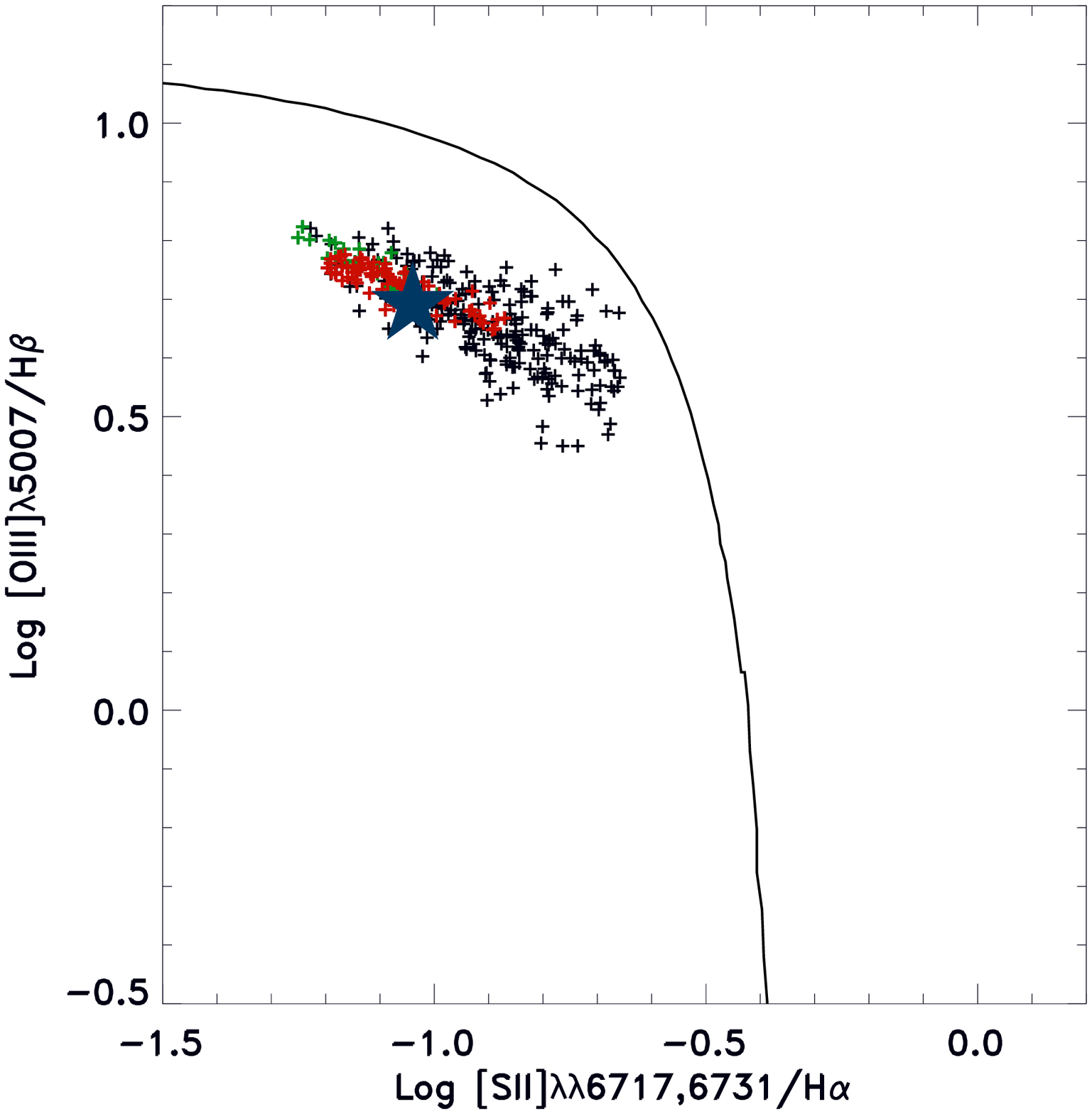}
\caption{[OIII]/H$\beta$ versus [SII]/H$\alpha$ diagnostic diagram for the original data points
presented in Figure \ref{ratio}a) and c). Red crosses correspond to region A, green crosses to
region B and the black crosses to the outer part of the galaxy. The
integrated value for the galaxy is represented by the
blue star. The solid line is the dividing line between HII galaxies and AGNs \citep{O6}.}
\label{diag}
\end{figure}

Figure \ref{diag} shows the  log[OIII]/H$\beta$ vs. log[SII]/H$\alpha$ diagnostic diagram, identifying
three different regions in UM 408 using the original data points (non-smoothed). Red and green crosses correspond to regions
A and B, respectively, and the black crosses correspond to the
outer parts of the galaxy. The integrated value that represents the sum over all spaxels in 
our field of view is given by the blue star symbol.
The solid line in Figure~\ref{diag} divides the diagram into two regions that correspond to line ratios explained solely by photoionization by massive stars and to line ratios where an additional source of excitation must be present, as in the case of Active Galactic Nuclei \citep[AGNs;][]{O6}. From the distribution of the data in the diagnostic diagram we conclude that the HII regions in UM 408 are produced by photoionization by massive stars.

\subsection{Density, temperature and oxygen abundance}\label{abundance}

\begin{figure*}[!hbt]
\centering
\includegraphics*[width=5cm]{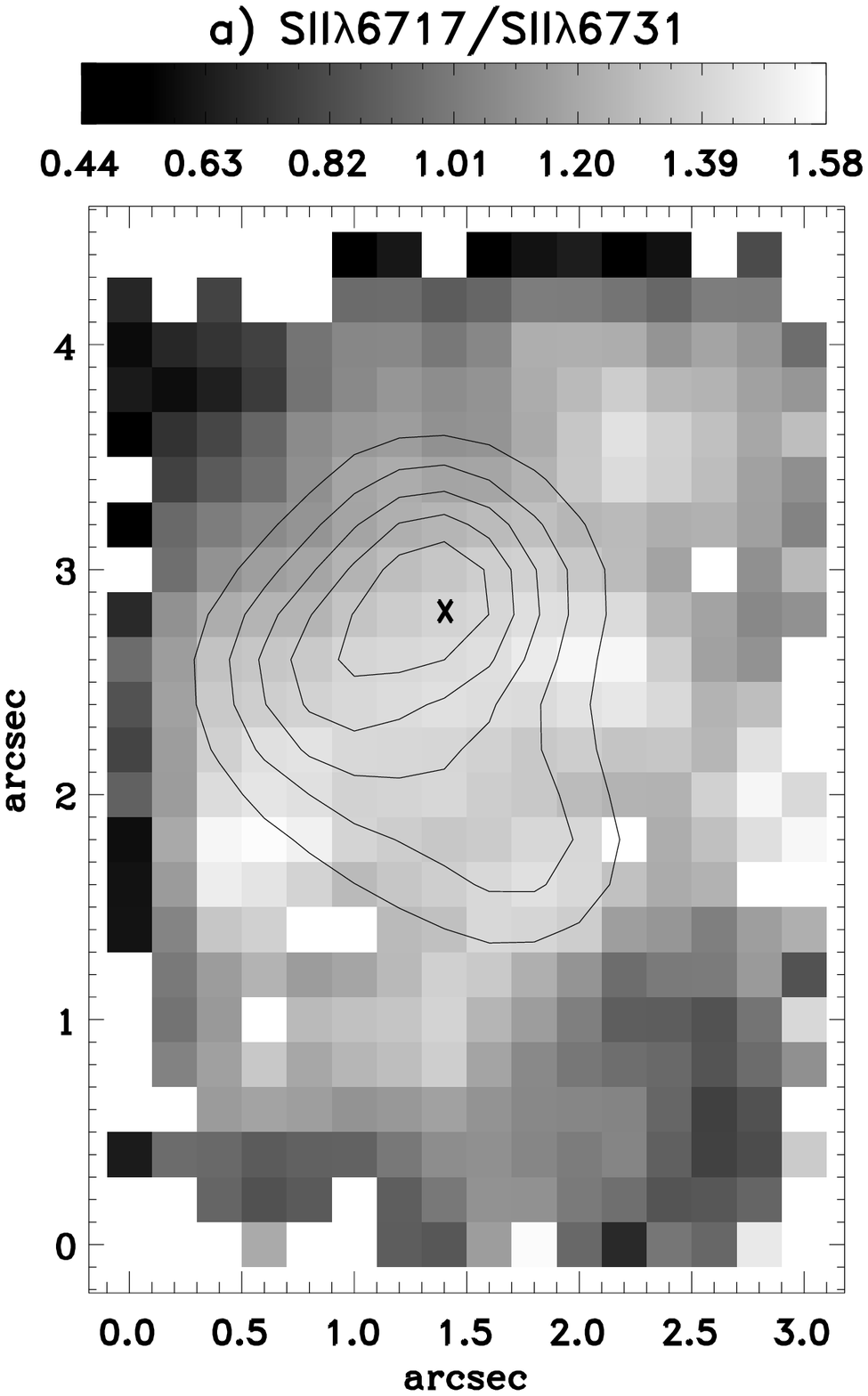}
\includegraphics*[width=5cm]{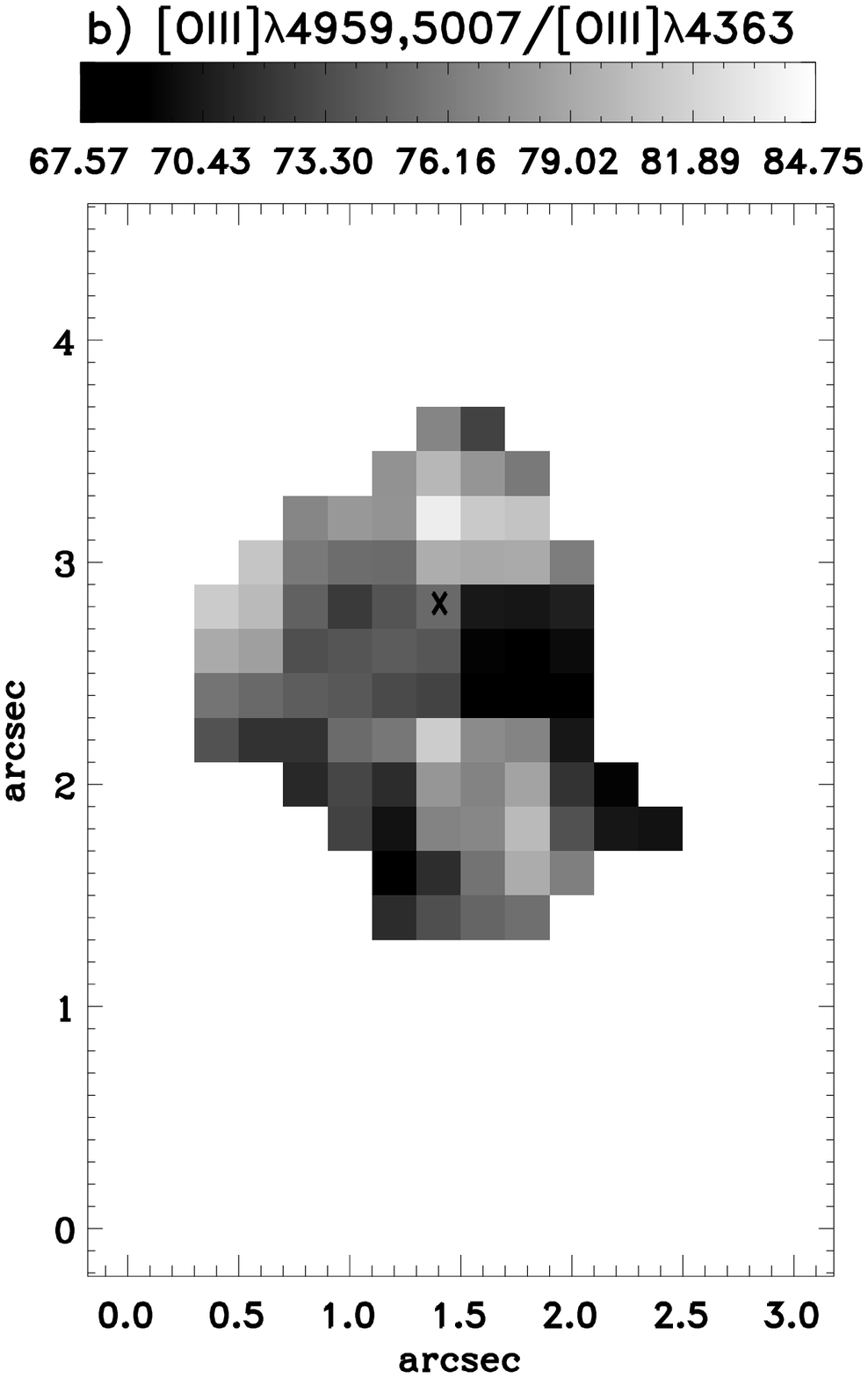}
\includegraphics*[width=5cm]{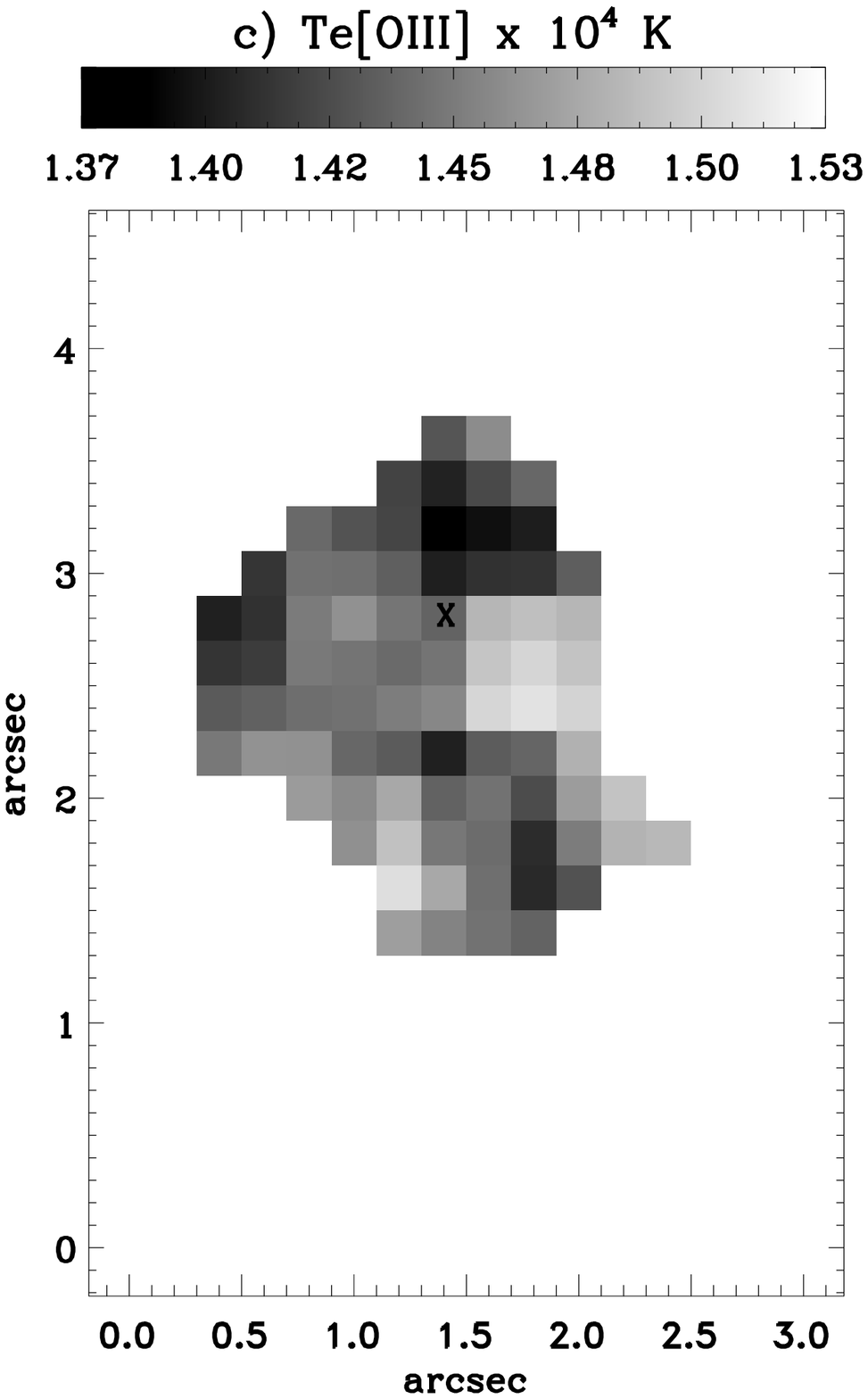}
\caption{(a) [SII]$\lambda$6717/[SII]$\lambda$6731 ratio map used to calculate
the electron density. (b) [OIII]$\lambda\lambda$4959,5007/[OIII]$\lambda$4363 emission line ratio,
which is used to calculate the electron temperature T$_{e}$(OIII). (c)
T$_{e}$(OIII) electron temperature in units of 10$^{4}$K. The maximum
H$\alpha$ emission is placed over region A and is indicated in the maps by an X symbol. H$\alpha$ contours are overplotted on the maps.}
\label{tempden}
\end{figure*}

The first step in the abundance derivation is to obtain the electron density and
temperature. The electron density is obtained using the emission line ratio
[SII]$\lambda$6717/[SII]$\lambda$6731, and the electron temperature from the
ratio [OIII]$\lambda\lambda$4959,5007/[OIII]$\lambda$4363.  In Figure
\ref{tempden}a we can see that the ratio [SII]$\lambda\lambda$6717,6731 is
typically greater than 1 which indicates a low density regime \citep{O6}, hence,
we assumed an electron density of n$_{e}\sim$100 cm$^{-3}$ for all apertures in
our calculations. To obtain the electron temperature T$_{e}$(OIII) we used the
five level atomic model FIVEL \citep{D87}, implemented under the IRAF STS
package NEBULAR. This program has been developed to obtain the physical
conditions in a low-density nebula, given appropriate emission line ratios.
Figure \ref{tempden}b shows the
[OIII]$\lambda\lambda$4959,5007/[OIII]$\lambda$4363 emission line ratio used to
calculate the electron temperature T$_{e}$(OIII), and Figure \ref{tempden}c
shows the resulting spatial distribution of electron temperature T$_{e}$(OIII).
The range of valid data points varies from 1.20 to  1.70 $\times$ 10$^{4}$ K
with a standard deviation of 0.1 $\times$ 10$^{4}$ K for the original data set
and from $\sim$1.37 to $\sim$1.53 $\times$ 10$^{4}$ K with a standard deviation
of 0.03 $\times$ 10$^{4}$ K for the smoothed data points. The total oxygen
abundances for each aperture are obtained assuming the contribution of O$^{+}$
and O$^{++}$ ions, thus we have that

\begin{equation}
\frac{O}{H}=\frac{O^{+}}{H^{+}} + \frac{O^{++}}{H^{+}},
\end{equation}

\noindent
where the T$_{e}$(OII) temperature and the O$^{+}$ and O$^{++}$ ions are obtained
assuming the expressions given by \cite{P92}.

\begin{figure*}[!htb]
\centering
\includegraphics*[width=8.0cm]{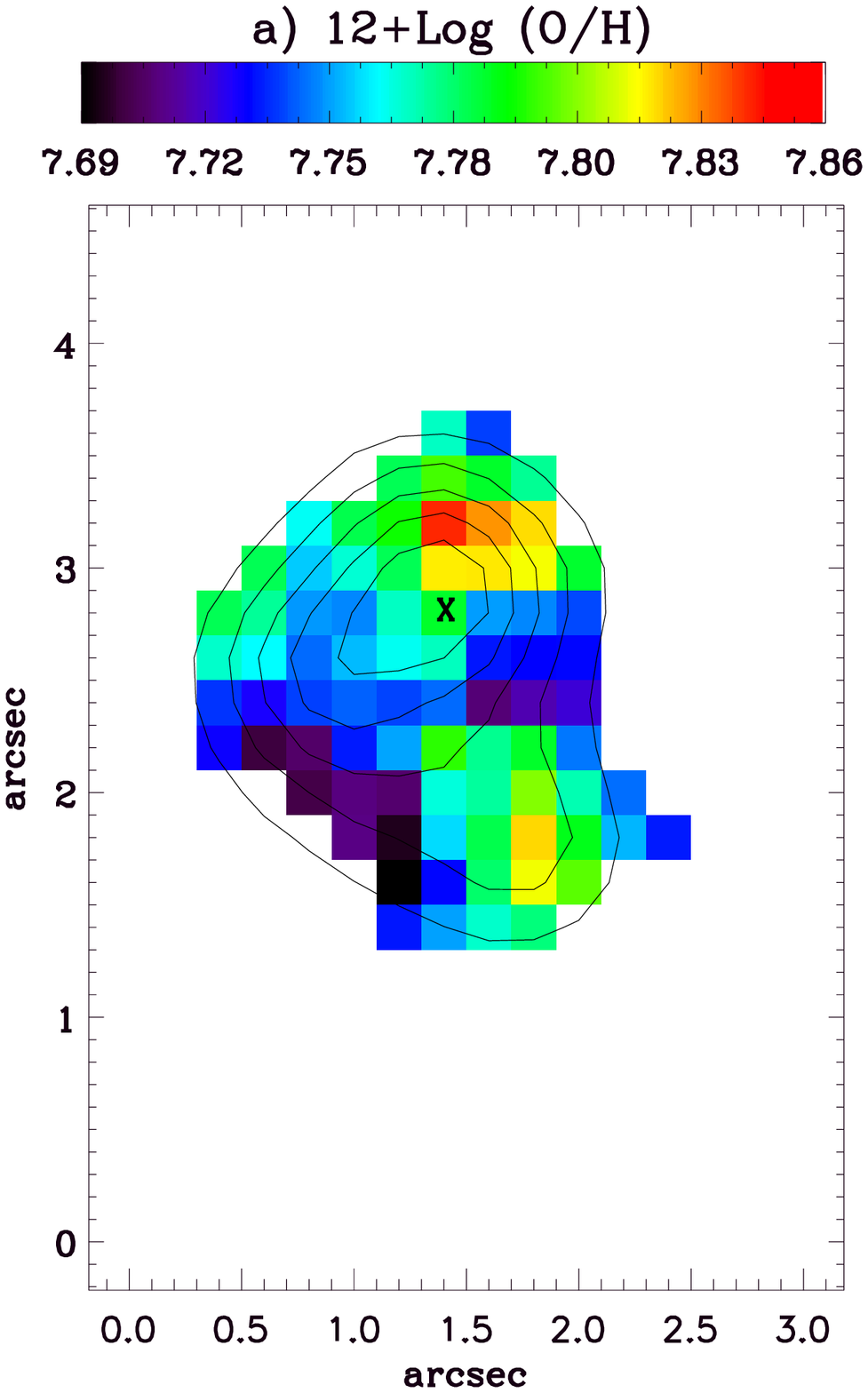}
\includegraphics*[width=8.0cm]{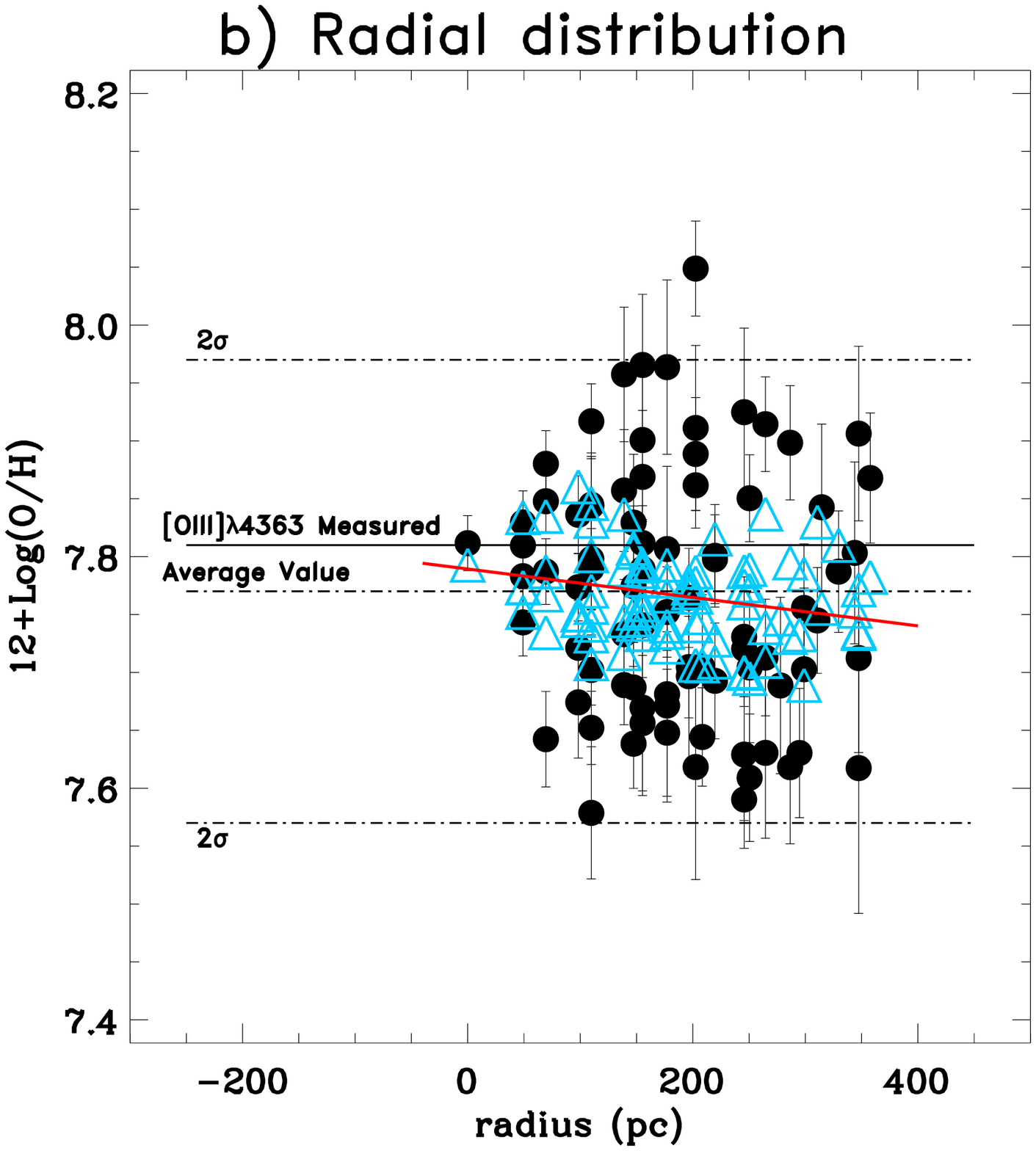}
\caption{Oxygen abundances [12+Log(O/H)]: (a) Spatial distribution. 
The isocontours display the H$\alpha$ emission. The maximum
H$\alpha$ emission is placed over region A and is indicated in the map by an X symbol. 
H$\alpha$ contours are overplotted on the maps. (b) Radial distribution.  
The central dotted line represents the 12+log(O/H) average value of 7.77, and 
the solid line the integrated value of 12+log(O/H)=7.81 in the region where the emission lines [OIII]$\lambda$4363 were measured.
The difference between the maximum and minimum values in the original data set (black dots) is equal to 0.47 dex. 
Light blue triangles represents the smoothed 12+Log(O/H) radial distribution. 
A least square fit using the original data points is indicated by the red line.}
\label{abun}
\end{figure*}

Figure \ref{abun}a shows the spatial distribution of the oxygen abundances over
the regions where the emission line [OIII]$\lambda$4363 was detected. The oxygen
abundance values in units of 12+log(O/H) range from 7.70$\pm$0.08 to
7.84$\pm$0.04 with a maximum error of $\sim$0.12 dex, an average value of 7.77
and a standard deviation of 0.10 dex. However, the smoothed map has values from
$\sim$7.69 to $\sim$7.86 with a standard deviation of $\sim$0.04 dex. Figure
\ref{abun}b shows the radial distribution of oxygen abundance with respect to
the H$\alpha$ peak emission within a region $\lesssim$ 0.4 kpc. Statistically
the bulk of the original data points are lying in a region of 2$\sigma$
dispersion ($\sigma$=0.1 dex) around the average value. The 2$\sigma$ dispersion
is represented by the dotted lines. A least fit square with a slope of
-0.14$\pm$0.06 dex kpc$^{-1}$ was found, using all original data points in
Figure \ref{abun}b. The error in the gradient is obtained directly from the
linear regression. In the same Figure we show the radial distribution of the
smoothed data points as light blue triangles. These results indicate that there
is no significant variations across the galaxy. At most, we can say that
there is a very marginal gradient of a decreasing abundance from the center
outwards, indicating that, on average, the highest abundance values are found
near the peak H$\alpha$ emission.

\subsection{Velocity field}

 The internal structure of HII galaxies and GHIIRs, as viewed in the ionized
gas, is directly associated with mechanisms of photoionization, magnetic field
induced turbulence, and feedback by the current episode of massive star
formation and evolution, i.e. radiative shocks, stellar and supernovae
driven-winds. All of it, under the influence of the gravitational potential of
the complex of gas and stars. The presence of expanding structures (shells,
loops and bubbles) is very common in GHIIRs \citep[e.g., the HST image on NGC
604,][for 30 Doradus]{T00,C94}. These structures have also been observed in the 
prototypical HII galaxies as II Zw 40 and their resulting
nested filaments have been explained by the effects of photoionization and stellar winds
in an inhomogeneous ISM \citep{TT06}.
UM 408 shows, in the H$\alpha$ emission line
map (Figure \ref{line1}) and in the acquisition broad-band image (Figure
\ref{image}), an outer regular shape with no signs of large-scale
disruption or a galactic wind \citep[see][and references therein for the case of
NGC 1569]{W07b}. 
The monochromatic maps, presented in this study, do not provide clear evidence of 
expanding structures, shell like features or filaments.

\begin{figure*}[!htb]
\centering
\includegraphics*[width=0.45\columnwidth]{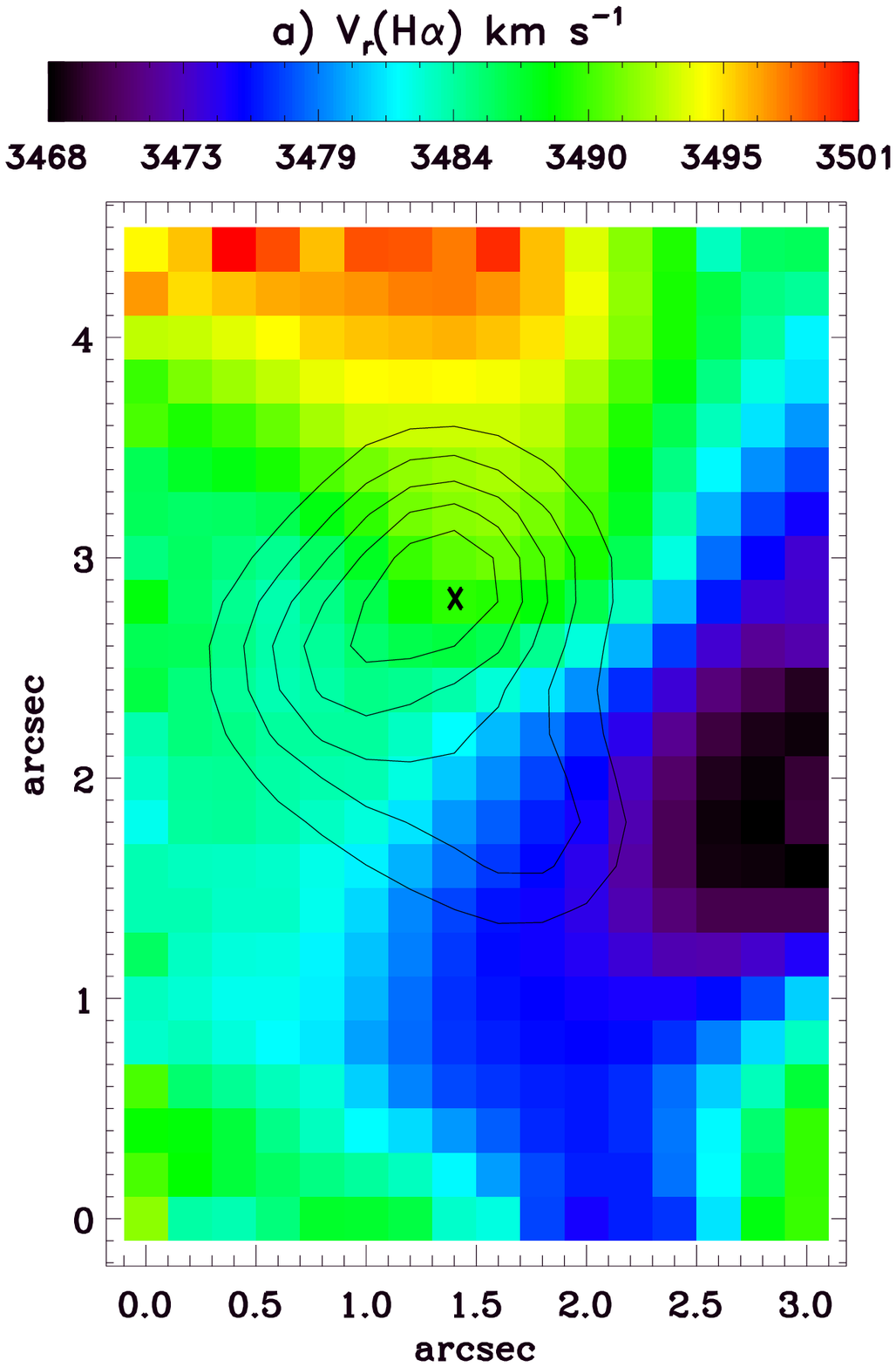}
\includegraphics*[width=0.45\columnwidth]{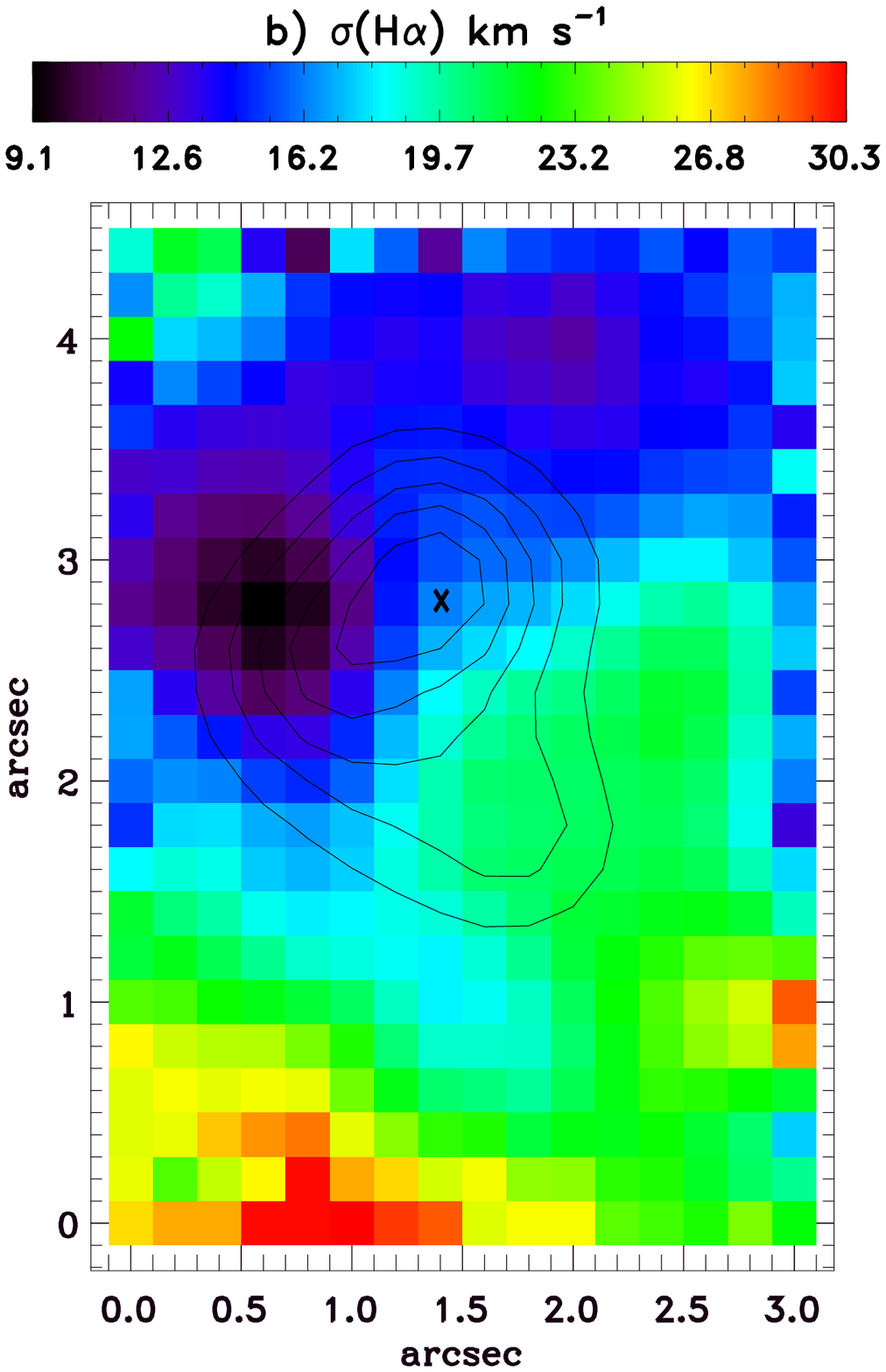}
\caption{(a) H$\alpha$ radial velocity in units of km s$^{-1}$.
(b) H$\alpha$ Velocity dispersion $\sigma$. The value showed in
this figure is obtained considering an instrumental $\sigma_{inst}$ 
and thermal $\sigma_{th}$ correction over the observed value $\sigma_{obs}$. The maximum
H$\alpha$ emission is placed over region A and is indicated in the maps by an X symbol.}
\label{dispersion}
\end{figure*}

In order to study the impact of the star formation on its
internal kinematic properties, we derived the spatial distribution of radial velocity v$_{r}$ (heliocentric) and velocity dispersion $\sigma$ (FWHM/2.355) by fitting a single Gaussian to
the H$\alpha$ emission line profiles. Figure~\ref{dispersion} shows the radial velocity
derived from the shifts of  the H$\alpha$ line peak, and
dispersion velocity maps calculated by using the following relation

\begin{equation}
\sigma^{2}(H\alpha)=\sigma^{2}_{obs}-\sigma^{2}_{inst}-\sigma^{2}_{th},
\label{eqdispersion}
\end{equation}

\noindent 
where $\sigma_{obs}$ is the observed H$\alpha$ velocity dispersion,
$\sigma_{inst}$ is the instrumental dispersion, and $\sigma_{th}$ is a
correction for thermal broadening ($\sqrt{kT_{e}/m_{H}} \approx$9.1 km s$^{-1}$
for hydrogen at a temperature of T$_{e}$=10$^{4}$ K). We have used a value of
33.4$\pm$1.0 km s$^{-1}$ for the instrumental dispersion by comparing the
nominal value (from Table~\ref{obslog}) with that measured directly from the
calibration lamps in the grim R600 near H$\alpha$, and assigning the error as
the rms of the average. 

The velocity field (Figure \ref{dispersion}a) shows an apparent systemic
motion where the east part of the galaxy (near knot B) is blueshifted, while the
west part (near knot A) is redshifted, with a relative motion of $\sim 10$ km
s$^{-1}$. This motion may contribute to broaden the integrated line profile
over the extent of the whole galaxy. However, it is not enough to explain the
total integrated supersonic line width observed of $\sigma = 18.6 \pm 1.5$ km
s$^{-1}$. 
Figure \ref{dispersion}b shows the velocity dispersion (line width)
map over the observed field, ranging values from $\sim 10$ km s$^{-1}$ to $\sim 30$ km s$^{-1}$.
The velocity dispersion varies little across the field, but some
points are worth noting. This map is characterized by a subsonic area near region A, while the highest 
values are located in the outer part of the field of view near region B.
The one pixel line width over knot A (the brightest
region) has a low $\sigma$ of 17.1$\pm 1.0$ km s$^{-1}$. This value is
identical to a synthetic aperture of 2.7\arcsec~over knot A. This synthetic
aperture was chosen to mimic the single fiber high dispersion observation of
\cite{B04} who find $\sigma$=19.6$\pm$0.4 km s$^{-1}$, that is consistent with our results.

\section{Discussion}\label{discussion}

\subsection{The integrated properties of UM 408}

We have measured the flux in two different apertures enclosing more than
$\sim$60$\%$ of the observed H$\alpha$ flux of the galaxy in order to study the
properties of the star formation regions detected in this galaxy. One aperture
for region A (48$\%$ of the measured H$\alpha$ flux) and another for region B
(13$\%$ of the measured H$\alpha$ flux). These regions have a diameters of 
about 1$\arcsec$.5 and 1$\arcsec$ that  correspond to
approximately 375 pc and 250 pc, respectively. Regions A and B are indicated in
the [OIII]$\lambda$5007 map of Figure \ref{line1}. These regions are formed by
the data points where the [OIII]$\lambda$4363 emission line is detected. To
obtain the integrated properties of the galaxy and compare with previous
observations \citep[e.g.][]{M94, P02} we derived the integrated spectra summing
over all pixels in our data cube. Table~\ref{inte-emission} summarizes the
resultant emission line values, obtained using the different synthetic
apertures, including the total [OIII]$\lambda$4363 aperture over the galaxy (in
regions where [OIII]$\lambda$4363 emission line is detected and measured). These measurements
are in reasonable agreement with those derived with long slit spectroscopy
\cite[e.g.,][]{T91,M94,P02} considering aperture differences and observational
uncertainties. A larger discrepancy however occurs between our observed
[OII]$\lambda\lambda$3726,29/H$\beta$ value of 0.52 and that of \cite{P02} of 1.45.
However, their value must be overestimated since, from their Figure 1, we
clearly see that H$\beta >$ [OII]$\lambda$3727.

The observed EW(H$\beta$) values in UM 408 are $\sim$102 $\rm \AA$ and $\sim$57
$\rm \AA$ for regions A and B and $\sim$67 $\rm \AA$ for the integrated galaxy,
respectively. The values for regions A and B have been calculated by taking the
average over all pixels in each region. Using these EW(H$\beta$) values we can
estimate the cluster ages by comparison with the values obtained from
STARBURST99 model \citep{L99}. The resulting ages assuming an instantaneous
burst and a metallicity of Z=0.004 for a Salpeter IMF with a mass limit of
1M$_{\odot}$ to 100M$_{\odot}$ are 4.83 Myrs and 5.00 Myrs for regions A and B,
respectively. These ages are consistent with both regions being formed
simultaneously. We found integrated luminosities (assuming a distance of D=46.8
Mpc; Pustilnik et al. 2002) of Log L(H$\alpha$)=39.15, 38.58 and 39.47 erg s$^{-1}$ for regions A and B
and for the integrated galaxy, respectively. These correspond to a star
formation rate of SFR=0.011, 0.003 and 0.024 M$_{\odot}$ yr$^{-1}$, and with a
number of ionizing photons log Q(H$^{0}$)=51.01, 50.45 and 51.34 photons
s$^{-1}$ for regions A, B and for the integrated galaxy, obtained from the
relationships given by \citet{K98}. The masses of these regions can be estimated
by scaling the number of ionizing photons with the models predicted for the
corresponding ages. The stellar masses are 5.17 $\times$ 10$^{4}$ M$_{\odot}$
and 2.29 $\times$ 10$^{4}$ M$_{\odot}$ for regions A and B, respectively. Table
\ref{integrated} shows the derived integrated properties of the galaxy and its
star forming regions.

Since its discovery, UM 408 was classified as a low abundance galaxy with an
integrated value of 12+Log(O/H)=7.63 for low resolution spectroscopy \citep{M94} and 7.93
for high resolution spectroscopy \citep{P02}. In this work, we found an
integrated abundance (summing over all spaxels in our field of view) of
12+Log(O/H)=7.87$\pm$0.05. The synthetic apertures previously mentioned, yield
an integrated oxygen abundance of 12+Log(O/H)=7.82$\pm$0.05 and 7.84$\pm$0.06
for regions A and B, respectively. Measuring the abundance in an aperture
equivalent with the area where the emission line [OIII]$\lambda$4363 was
measured we obtain a value of 7.81$\pm$0.05, the same value obtained in the peak
of the H$\alpha$ emission (see Figure \ref{abun}b). This result reflects the
fact that the integrated values are light weighted and dominated by the galaxy
peak emission.

Finally, despite the fact that a detailed kinematic analysis is beyond the scope
of the present paper, our results confirm that the core dominates the kinematic
information of the star forming region, and present a low $\sigma$ value, as
found by other works \citep[e.g.,][and references therein]{tel01}. 
Therefore, a single Gaussian fit to the line profile of any aperture
encompassing the brightest knot will measure a representative line width of the
dominant supersonic motions, somewhat deconvolved by the effects of stellar and
SN mechanical energy input. The effects of stellar or SN feedback will
contribute to the broadening of the line profiles, but will only be detectable
in the lower density regions at low intensities. One means to recognize these
effects is through the identification of inclined bands in  the diagnostic
diagrams of intensity vs. $\sigma$ as proposed by \cite{mun96} and \cite{Y96}. 
However, the compactness of UM 408 and the small range of reliable line widths
make the use of the diagnostic diagrams difficult to interpret in this case.
The H$\alpha$ line profile is symmetric and well
represented by a single Gaussian, and does not show prominent low intensity
wings in either apertures. An attempt to deblend a possible additional component
to this line profile produced only an upper limit of $\sigma_{broad} \leq 100$
km s$^{-1}$ with $F(\alpha)_{broad}/F(\alpha)_{narrow} \leq 3\%$.

\subsection{Distribution of the oxygen abundance and their comparison with other dwarf galaxies}

The uniform behavior of O/H abundance in scales of hundreds of pc in low mass
galaxies as in UM 408 is not without precedent, and is comparable with the
observed variation in some dwarf galaxies by other authors. In NGC 4214 (Dwarf
Irregular/Wolf-Rayet galaxy) \citet{K96} showed that there is no localized O, N
or He abundance gradients, but they found, near the youngest region, an oxygen
overabundance of 0.095$\pm$0.019 dex. \citet{K97b}, using long slit spectroscopy
to study the interstellar medium of the dwarf irregular galaxy NGC 1569, failed
to  find evidence for O/H gradient from the recent star formation activity.
\citet{Lee06} studied the local group dwarf irregular galaxy NGC 6822 and
measured a difference of $\Delta$(O/H)=0.53 dex between the maximum
(8.43$\pm$0.2) and minimum value (7.90$\pm$0.1), with a mean oxygen abundance of
12+log(O/H)=8.11$\pm$0.1 using only the HII regions where the
[OIII]$\lambda$4363 emission lines were detected. A gradient of -0.14$\pm$0.07
dex kpc$^{-1}$ was measured by Lee et al. in a radius $\lesssim$1.4 kpc, and a
slope of -0.16 dex kpc$^{-1}$ was obtained if measurements from the literature
were included, showing the existence of a possible radial gradient in oxygen
abundance. \citet{I06} using VLT/GIRAFFE in the ARGUS mode, observed the dwarf
galaxy SBS 0335-052E, one of the most metal poor galaxies with an integrated
oxygen abundance of 12+log(O/H)=7.30 \citep{M92}. The spatial distribution of
the oxygen abundance in the ISM of SBS 0335-052E vary from 7.00$\pm$0.08 to
7.42$\pm$0.02, representing a difference of $\Delta$(O/H)=0.42 dex. The maximum
value of oxygen abundances in this galaxy does not correspond to the position of
one of  the identified SSCs, but Izotov et al. find a slight trend of a
decreasing  abundance, interpreted as possible self-enrichment. However, they
argue that the errors in the calculated oxygen abundances and T$_{e}$, n$_{e}$,
etc, not considering instrumental and observational uncertainties, lead to
consider the oxygen abundance variations in SBS 0335-052E may not be
statistically significant. More recently, using IFU-PMAS observations,
\citet{K08} studied the spatial distribution of some metals over the galaxy
IIZw70. In this study they find a difference of $\bigtriangleup$(O/H)=0.40 dex 
between the maximum (8.05$\pm$0.06) and minimum (7.65$\pm$0.06) values of oxygen
abundance. In the case of UM 408, the bulk of our observed data points are
distributed in a region of $\pm$2$\sigma$ dispersion around the average value with a
difference between the maximum and minimum values of 0.47 dex. This is,
somewhat, in good agreement with previous results in other dwarf galaxies, using
in some cases different techniques. We note that the gradient found in our work,
calculated in a spatial scale of hundreds of pc, is similar to the gradients
calculated in other studies in scales of kpc. Finally, the two
giant regions A and B show a difference of oxygen abundance of only $\bigtriangleup$(O/H)=0.02 dex, 
which indicates that these regions show identical chemical properties within the errors.  
Given the ages ($\sim$5 Myr) and stellar masses ($\sim$10$^{4}$M$_{\odot}$) of these regions, 
we expect that hundreds of SNs have exploded, producing eventually localized enrichments. 
But the absence of chemical overabundances in the ISM of UM 408 and in the dwarf 
galaxies studied in the literature lead to conclude that the population of young clusters 
have not produce preferentially localized overabundances.

These results are compatible with the interpretation that the newly synthesized
metals from the current star formation episode may not be in the warm phase of
the ISM, and thus are not observed at optical wavelengths. The metals that are
observed, however, ought to be from previous events, and may be well mixed and
homogeneously distributed over the whole extent of this low mass and compact
galaxy. Note that in UM 408, there is no bar induced rotation, or shear, to
produce the metal dispersal and mixing as in more massive galaxies
\citep[e.g.,][]{R95}, and thus there must be another mechanism responsible for
the large-scale dispersal and mixing with the ISM. One such hydrodynamic
mechanism for the transport and mixing of the metals produced by compact bursts
of star formation into the ISM has been proposed by \citet{T96}. This model
predicts that the energy injected by core-collapse supernovae change the
physical conditions (density and temperature) of the interstellar medium, while
undergoing a long excursion into the galactic volume. The injected matter is
first thermalized, near the starburst, as it goes through a reverse shock. This
generates the giant pressure that allows for the built up of superbubbles, kpc
scale structures that may grow for up to 50 Myr. Afterwards, once the SN II
phase is over, the giant superbubble gas  begins to cool down by radiation. This
 happens first within the parcels of gas that present the largest densities,
promoting their thermal instability. The sudden loss of pressure within densest 
parcels of gas would immediately trigger the appearance of re-pressurizing
shocks emanating from the low density hot gas. The process leads to a plethora
of dense condensations that inevitably will fall and settle within the main body
of the galaxy. The process of cooling and dispersal onto the main body of the
galaxy will occur within a few 10$^8$ yr. For a complete mixing, the model
assumes that a further star formation episode has to occur and through
photoionization the metal enrich condensations will expand and be fully mixed
with the ISM \citep[see also][]{S07}. This scenario seems plausible and
compatible with our analysis for the case of UM 408.

\section{Conclusions}\label{conclusions}

We used GMOS-IFU spectrum of the compact HII galaxy UM 408,
in order to derive the spatial distribution of emission line ratios, extinction 
c(H$\beta$), radial velocity v$_{r}$, velocity dispersion $\sigma$, electron temperature and oxygen
abundance (O/H) as well the integrated properties over an extended region of
3$\arcsec\times$4$\arcsec$.4 equivalent to 750 $\times$ 1100 pc in the central part of the galaxy. 
Below, we summarize our results and conclusions.

\begin{enumerate}

\item The observed region of the galaxy includes two giant HII regions 
not resolved in previous studied, here labelled A and B. The sizes of these ionized regions are $\sim$375 and $\sim$250 pc for regions A and B, respectively. 
Another region labelled C was found to the East of region B. We do not observe large scale
structures as in other GHIIRs and HII galaxies in the monochromatic maps.

\item The distribution of dust content in the galaxy was derived from  
the Balmer line decrement (H$\alpha$/H$\beta$). The c(H$\beta$) distribution concentrates 
the highest values close but not coincident with the peak of H$\alpha$ emission 
in each ionized cluster (A and B). The dust seems to be displaced from 
the star formation regions by the action of the star cluster winds. 

\item We used [OIII]/H$\beta$ and [SII]/H$\alpha$ ratios to investigate
the possible excitation mechanism over the ISM of the galaxy. Comparing 
the data points in the diagnostic diagram with the theoretical model 
given by \cite{O6}, we found that all measured H$\alpha$ flux arises from gas 
photoionized by massive stars

\item The velocity field (v$_{r}$) shows an apparent systemic
motion where the east part of the galaxy (near knot B) is blueshifted, while the
west part (near knot A) is redshifted, with a relative motion of $\sim 10$ km s$^{-1}$ 
and a difference between the maximum a minimum values of $\sim$33 km s$^{-1}$. 
The velocity dispersion map shows supersonic values, typical for 
extragalactic HII regions, ranging values from $\sim 10$ km s$^{-1}$ to $\sim 30$ km s$^{-1}$.

\item The ages of the two giant regions detected in this study, were estimated from their EW(H$\beta$) 
and the STARBURST99 models, suggesting that they are coeval events of $\sim$5 Myr with 
stellar masses of $\sim$10$^{4}$ M$_{\odot}$. We see a marginal difference in the 
integrated oxygen abundance between these regions of $\bigtriangleup$(O/H)=0.02 dex.

\item Finally, as found in other nearby dwarf galaxies, we do not observe a gradient 
in oxygen abundance across the compact galaxy UM 408. The bulk of the observed data points 
are lying in a region of $\pm$2$\sigma$ dispersion ($\sigma$=0.1 dex); therefore, the new metals formed
in the current star formation episodes are not observed and reside probably in
the hot gas phase (T$\sim$10$^{7}$ K), whereas the metals from previous star formation events are well mixed and homogeneously distributed through the whole extent of the galaxy.

\end{enumerate}

All results presented here are suggestive that UM 408 is an unevolved low metallicity dwarf galaxy, 
undergoing a simultaneous episode of star formation over the whole optically observed extension, and
it is a genuine example of the simplest starbursts occurring in galactic scale, possibly mimicking the properties one expects for young galaxies at high redshift.

\section{Acknowledgments}
We would like to thank the anonymous referee for his/her comments and
suggestions which substantially improved the paper. PL acknowledges support from
Faperj, Brazil for his studentship and IAC Spain where part of this work was
done. This work has been partly funded by ESTALLIDOS (see
http://www.iac.es/project/GEFE/estallidos) project AYA2007-67965-C03-01. ET
acknowledges his US Gemini Fellowship by AURA, and GTT acknowledges a research
grant 60333 from CONACYT Mexico. Based on observations obtained at the Gemini Observatory, which is operated by the
Association of Universities for Research in Astronomy, Inc., under a cooperative agreement
with the NSF on behalf of the Gemini partnership: the National Science Foundation (United
States), the Science and Technology Facilities Council (United Kingdom), the
National Research Council (Canada), CONICYT (Chile), the Australian Research Council
(Australia), Minist\'erio da Ci\^encia e Tecnologia (Brazil) and Ministerio de Ciencia,
Tecnolog\'{\i}a e Innovaci\'on Productiva  (Argentina); Gemini program ID:
GS-2004B-Q-59. This research has made use of the NASA/IPAC Extragalactic
Database (NED) which is operated by the Jet Propulsion laboratory, California
Institute of technology, under contract with the National Aeronautics and Space
Administration.

\clearpage

\begin{deluxetable}{ccccccc}\label{obslog}
\tabletypesize{\scriptsize}
\tablecaption{Observational setup}
\tablenum{1}
\tablecolumns{6}
\tablewidth{0pt}
\tablehead{
\colhead{Grating} &
\colhead{Central Wavelength}&
\colhead{Airmass} &
\colhead{Exposure Time}&
\colhead{Resolution}&
\colhead{Dispersion}&
\colhead{Seeing}
\\
\colhead{}&
\colhead{(\AA)}&
\colhead{}&
\colhead{(seconds)}&
\colhead{(\AA)}&
\colhead{(\AA/pixel)}&
\colhead{(\arcsec)}
\\
\colhead{(1)}&
\colhead{(2)}&
\colhead{(3)}&
\colhead{(4)}&
\colhead{(5)}&
\colhead{(6)}&
\colhead{(7)}
}
\startdata
B600 & 4300 & 1.24 & 1$\times$1800 & 1.98 & 0.45 & 0.8\\
R600 & 5850 & 1.47 & 2$\times$2000 & 1.67 & 0.47 & 0.9\\
\enddata
\end{deluxetable}

\clearpage

\begin{deluxetable}{ccccc}
\tabletypesize{\scriptsize}
\tablecaption{Observed emission line fluxes for the integrated galaxy and regions A and B \label{inte-emission}. Fluxes normalized to F(H$\beta$).}
\tablenum{2}
\tablecolumns{4}
\tablewidth{0pt}
\tablehead{	
\colhead{}&
\colhead{Integrated galaxy}&
\colhead{[OIII]$\lambda$4363 measured}&
\colhead{Region A}&
\colhead{Region B}
\\
\colhead{}&
\colhead{(1)}&
\colhead{(2)}&
\colhead{(3)}&
\colhead{(4)}
}
\startdata
$\left[SII\right]\lambda$6731     & 0.24$\pm$0.01 & 0.20$\pm$0.03 &  0.20$\pm$0.03   & 0.20$\pm$0.04\\
$\left[SII\right]\lambda$6717     & 0.32$\pm$0.01 & 0.28$\pm$0.04 &  0.27$\pm$0.03   & 0.27$\pm$0.04\\
$\left[NII\right]\lambda$6584     & 0.13$\pm$0.01 & 0.11$\pm$0.03 &  0.11$\pm$0.02   & 0.10$\pm$0.03\\
H$\alpha$                         & 5.94$\pm$0.07 & 6.07$\pm$0.29 &  6.14$\pm$0.28   & 5.87$\pm$0.28\\
$\left[OIII\right]\lambda$5007    & 5.24$\pm$0.21 & 5.33$\pm$0.30 &  5.30$\pm$0.25   & 5.60$\pm$0.36\\
$\left[OIII\right]\lambda$4959    & 1.71$\pm$0.18 & 1.72$\pm$0.11 &  1.71$\pm$0.11   & 1.80$\pm$0.11\\
$\left[OIII\right]\lambda$4363    & 0.06$\pm$0.01 & 0.07$\pm$0.01 &  0.07$\pm$0.01   & 0.07$\pm$0.02\\
H$\gamma$                         & 0.29$\pm$0.01 & 0.31$\pm$0.04 &  0.31$\pm$0.04   & 0.36$\pm$0.05\\
$\left[OII\right]\lambda\lambda$3726,29& 0.52$\pm$0.17 & 0.30$\pm$0.05 &  0.32$\pm$0.04   & 0.27$\pm$0.05\\
\hline
F(H$\beta$)                       & 39.70$\pm$0.20 & 24.93$\pm$0.86 & 18.63 $\pm$0.62  & 5.09$\pm$0.17\\
\enddata
\tablecomments{Fluxes normalized to F(H$\beta$) in units of $\times$10$^{-16}$ erg s$^{-1}$ cm$^{-2}$. Column (1): Integrated galaxy correspond with the values obtained from the sum of all spaxels in the data cube. Column (2): [OIII]$\lambda$4363 measured correspond with sum of all apertures in the maps where [OIII]$\lambda$4363 was measured. Columns (3) and (4): Regions defined in Figure \ref{line1}. }
\end{deluxetable}

\clearpage

\begin{deluxetable}{ccccc}
\tabletypesize{\scriptsize}
\tablecaption{Integrated properties of the observed area of UM 408. \label{integrated}}
\tablenum{3}
\tablecolumns{4}
\tablewidth{0pt}
\tablehead{
\colhead{}&
\colhead{Integrated galaxy}&
\colhead{[OIII]$\lambda$4363 measured}&
\colhead{Region A}&
\colhead{Region B}
\\
\colhead{}&
\colhead{(1)}&
\colhead{(2)}&
\colhead{(3)}&
\colhead{(4)}
}
\startdata
c(H$\beta$)                                & 0.93$\pm$0.02 & 0.96$\pm$0.06      & 0.97$\pm$0.06& 0.91$\pm$0.06 \\
Te(OIII) $\times$ 10$^{4}$ (K)             & 1.40$\pm$0.08 & 1.45$\pm$0.14      & 1.45$\pm$0.12& 1.44$\pm$0.16 \\
Te(OII)  $\times$ 10$^{4}$ (K)             & 1.32$\pm$0.11 & 1.34$\pm$0.18      & 1.34$\pm$0.16& 1.34$\pm$0.21 \\
n$_{e}$ (cm$^{-3}$)                        & $\sim$100 &$\sim$100           &  $\sim$100   &  $\sim$100 \\
log$\left[OIII\right]\lambda5007/$H$\beta$ & 0.69$\pm$0.02 &  0.70 $\pm$0.02    & 0.69$\pm$0.02&  0.72$\pm$0.03\\
$\left[SII\right]$/H$\alpha$               & 0.09$\pm$0.01 &  0.08$\pm$0.01     & 0.07$\pm$0.01&0.08$\pm$0.02\\
O$^{+}$/H$^{+} \times$10$^{5}$             & 1.16$\pm$0.26 & 0.64$\pm$0.09      &0.70$\pm$0.09& 0.57$\pm$0.10\\ 
O$^{++}$/H$^{+}\times$10$^{5}$             & 6.35$\pm$0.58 & 5.88$\pm$0.68      &5.85$\pm$0.60&6.31$\pm$0.82 \\
12+log(O/H)                                & 7.87$\pm$0.05 &  7.81$\pm$0.05     & 7.82$\pm$0.05& 7.84$\pm$0.06\\
\hline
log L(H$\alpha$) (erg s$^{-1}$)           & 39.47 & 39.27     &    39.15     & 38.58 \\
log (number of photons) (photons s$^{-1}$)$^{\dagger}$& 51.34 & 51.14     &  51.01     & 50.45\\
SFR(M$_{\odot}$ yr$^{-1}$)$^{\dagger}$     & 0.024 & 0.015     &   0.011     &  0.003 \\
EW(H$\beta$) ($\rm \AA$)                   & 62 & 89        &   102      & 57 \\
Age (Myr)$^{\dagger\dagger}$               & \nodata & 4.73      &    4.83     & 5.00 \\
\enddata
\tablecomments{$^{\dagger}$ Number of photons and SFR were obtained using the relations given by \citet{K98}. $^{\dagger\dagger}$ Ages were obtained using the STARBURST99 predictions model, assuming an instantaneous star formation and a Salpeter IMF from 1 to 100 M$_{\odot}$.}
\end{deluxetable}

\end{document}